\documentclass[useAMS,usenatbib,usegraphicx]{mn2e}

\def\aap{A\&A}
\def\apj{ApJ}
\def\apjs{ApJS}
\def\apjl{ApJL}
\def\mnras{MNRAS}

\newcommand{\cl}{C_{\ell}}

\newcommand{\beal}{\begin{align}}
\newcommand{\id}{{\,\rm d}}
\newcommand{\Abst}[1]{\,#1}
\newcommand{\pot}[2]{#1 \times 10^{#2}}
\newcommand{\lesssim}{\mathrel{\hbox{\rlap{\hbox{\lower4pt\hbox{$\sim$}}}\hbox{$<$}}}}
\newcommand{\gtrsim}{\mathrel{\hbox{\rlap{\hbox{\lower4pt\hbox{$\sim$}}}\hbox{$>$}}}} 
\newcommand{\nmax}{n_{\rm max}} 
\newcommand{\nsplit}{n_{\rm sp}} 
 
\newcommand{\zem}{z_{\rm em}} 
\newcommand{\kB}{k}

\newcommand{\Tg}{T_{\gamma}}
\newcommand{\Te}{T_{\rm e}}

\usepackage{amsmath}

\title[Lines in the CMB Frequency Spectrum from Recombination]{Lines in the 
  Cosmic Microwave Background Spectrum from the Epoch of
  Cosmological Hydrogen Recombination}

\author[J.~A. Rubi\~no-Mart\'{\i}n, J. Chluba \& R.A. Sunyaev]{
J.~A. Rubi\~no-Mart\'{\i}n$^{1,2}$\thanks{E-mail: 
jose.alberto.rubino@iac.es}, 
J. Chluba$^2$\thanks{E-mail: jchluba@mpa-garching.mpg.de} 
and R.A. Sunyaev$^{2,3}$\\
$^{1}$ Instituto de Astrof\'{\i}sica de Canarias, C/V\'{\i}a L\'actea s/n, 
    E-38200 Tenerife, Spain\\
$^{2}$ Max-Planck Institut f\"ur Astrophysik, Karl-Schwarzschild-Str. 1,
D-85740 Garching, Germany\\
$^{3}$ Space Research Institute (IKI), Russian Academy of Sciences,
Moscow, Russia }

\begin{document}

\date{Received **insert**; Accepted **insert**}

\pagerange{\pageref{firstpage}--\pageref{lastpage}}
\pubyear{}

\maketitle

\label{firstpage}

\begin{abstract}
  We compute the spectral distortions of the Cosmic Microwave
    Background (CMB) arising during the epoch of cosmological hydrogen
    recombination within the standard cosmological (concordance) model for
    frequencies in the range $1\,{\rm GHz}-3500\,$GHz.
    We follow the evolution of the populations of the hydrogen levels
    including states up to principle quantum number $n=30$ in the redshift
    range $500\leq z\leq 3500$.
  All angular momentum sub-states are treated individually, resulting in a
  total number of $465$ hydrogen levels. The evolution of the matter
  temperature and the fraction of electrons coming from HeII are also 
  included.
We present a detailed discussion of the distortions arising from the
main dipolar transitions, e.g. Lyman and Balmer series, as well as the
emission due to the two-photon decay of the hydrogen 2s level.
  Furthermore, we investigate the robusteness of the results against
  changes in the number of shells considered.
The resulting spectral distortions have a characteristic oscillatory
behaviour, which might allow experimentalists to separate them from
other backgrounds. The relative distortion of the spectrum exceeds a
value of $10^{-7}$ at wavelengths longer than $21\,$cm.
Our results also show the importance of detailed 
  follow-up of the angular momentum sub-states, and their 
  effect on the amplitude of the lines.
  The effect on the residual electron fraction is only moderate, and
  mainly occurs at low redshifts. The CMB angular power spectrum is
  changed by less than 1\%.
Finally, our computations show that if the primordial 
radiation field is described by a pure blackbody, then 
there is no significant emission
from any hydrogen transition at redshifts greater than $z \sim 2000$.
This is in contrast to some earlier works, where the existence of a
  `pre-recombination' peak was claimed.
\end{abstract}


%
\section{Introduction}

In the last twenty years, many experiments have been devoted to
the study of the angular fluctuations of the  
Cosmic Microwave Background (CMB) temperature and polarization. Nowadays, 
the results of the WMAP 
satellite 
\citep{2003ApJS..148..175S,2006astro.ph..3449S} in conjunction with other
datasets (observations of type Ia supernovae or galaxy surveys)  
give us the most precise
determinations of the cosmological parameters describing our Universe, 
and constitute one of the most important confirmations of 
the hot Big Bang theory.
One of the predictions of this theory is the existence 
of small spectral distortions of the CMB blackbody spectrum
arising as a consequence of the non-equilibrium conditions occurring
during the epoch of the primordial hydrogen 
recombination at redshift $z\sim 1100$
\citep{ZeldKurtSun1968orig,1968ApJ...153....1P}. 
These authors showed that,
because in the Wien part of the CMB spectrum 
the brightness is extremely low, 
the strongest distortions are connected with the
Ly$\alpha$ line (which today should be found at $\lambda \sim 170~\mu$m), 
and with the 2s level two-photon decay emission (which
should peak around $\lambda \sim 200~\mu$m).
There should also be similar signatures arising from the
recombination of helium (which occurs around $z\sim 6000$ for HeIII
and $z\sim 2500$ for HeII), but these appear at longer wavelengths
than the hydrogen Ly$\alpha$ line
\citep[e.g.][]{1997AstL...23..565D,Wong2005} and they lie one order of
magnitude below the hydrogen spectrum, just due to the lower abundance
of helium.

\citet{Dubrovich1975} proposed to look for the spectral distortions
due to transitions between higher levels of the hydrogen atom 
(Balmer and higher series), and after
this, many papers were devoted  to the 
problem of the formation of the hydrogen 
recombination lines in the early Universe   
\citep{Liubarskii1983, RybickiDell93, 
  1995A&A...302..635D, 1997AstL...23..565D, Boschan1998, Burgin2003, 
  2004AstL...30..509D, Kholu2005, Wong2005}.
Although there were contradictory results in the amplitudes and shapes
of the features, these computations showed that 
the amplitude of the spectral distortions arising from 
transitions of the higher series is very small, with 
typical values around $\sim10^{-6}-10^{-8}$. 
The important point for us is that these distortions should
have survived until today and might become observable in the near future.

The measurements from the Far-InfraRed Absolute Spectrophotometer
\citep[FIRAS, e.g. ][]{1996ApJ...473..576F,Fixsen2002} on
board the COBE satellite indicate that the CMB frequency spectrum is
described by a $T_0 = 2.725\pm0.001$~K blackbody spectrum with high
precision.  The limits on the Bose-Einstein and Compton distortions
are $|\mu| < 9 \times 10^{-5}$ (95\% CL) and $|y| < 15 \times 10^{-6}$
(95\% CL), respectively.
Moreover, \citet{Fixsen2002} point out that recent
technological progress permits one to improve the limits by nearly
$2$ orders of magnitude.  Therefore, we are approaching the
sensitivities required for detecting global features in the
spectrum, so detecting individual ones could be possible
in the next future. Therefore, precise computations of the shapes,
positions and amplitudes of these lines are needed.
We also mention that there have been attempts to measure the 
CMB spectrum at centimeter wavelengths using the ARCADE experiment 
\citep{2004ApJS..154..493K, 2004ApJ...612...86F}. 

In most of the existing calculations simplifying assumptions 
have been used in order to make the problem computationally tractable. 
For the study of the Ly$\alpha$ line and the 2s two-photon decay emission, 
the effective three-level atom is usually adopted 
\citep[e.g.][]{ZeldKurtSun1968orig,1968ApJ...153....1P,Boschan1998,Wong2005}. 
%
For transitions in the Balmer and higher series, two assumptions are
usually adopted \citep[e.g.][]{Burgin2003,2004AstL...30..509D,Kholu2005}. 
First, a quasi-static evolution of the populations  
for all levels above $n=2$ is assumed, 
which reduces the problem of solving a system of
stiff coupled ordinary differential equations to 
an algebraic system. And second, given that one of the
main difficulties is treating {\it all} the energetically 
degenerate angular momentum 
sublevels within each shell, only the 
total population of a given shell is computed and
statistical equilibrium\footnote{This assumption implies that the population 
of a level $(n,l)$ is given by 
$N_{nl}=(2l+1) N_n/n^2$, where $N_n$ is the 
total population of the shell with principle quantum number $n$}
(SE) among the sublevels is assumed. 
To date, the only attempt to avoid these two assumptions was made by
\citet{RybickiDell93}, who considered the case of 
10-shells. However, their results were not
conclusive and were affected by numerical uncertainties. 
In particular until now the question whether the assumption 
of SE is justified has not been adressed satisfactory.

It is important to mention that all the codes which compute the
power spectrum of the angular fluctuations of the CMB make use
of the recombination history as obtained with the {\sc Recfast} code
\citep{Seager1999ApJ}. These calculations are
based on a $300$-level hydrogen atom 
in which SE among the $l$ sublevels was assumed 
\citep{Seager2000ApJS}. Therefore, it would also be important
to understand the validity of this approximation and its effects on the recombination history.

%
In our paper, we have obtained the spectral distortions from 
{\it all lines} of hydrogen 
including $30$-shells, following the evolution of 
{\it all angular momentum sub-states separately}.
Since we are until now not including collisions, from an estimate
following \citet{1964MNRAS.127..165P} this seems a natural choice in
order to keep the calculations simple.  Our Universe has extremely
high entropy, i.e. photons outnumber the baryons by a factor of $\sim
1.6 \times 10^{9}$. This is why collisions in general are {\it not}
important. However, those collisions connecting different,
energetically degenerate angular momentum substates within a given
shell should play a role for sufficiently high shells, but
the results presented here should remain unaffected.
A detailed discussion of this subject will be done in 
a subsequent paper (Chluba et al. 2006, in preparation).

Our computations have permitted us to show for the first time
that the deviations in the populations of the angular momentum
substates within a given shell from SE are important when studying the
spectral distortions to the cosmological blackbody,
and that the impact on the residual ionization fraction of the
hydrogen atoms after recombination is of the order of a few percent.
However, this change is sufficiently important to produce
modifications of the order of few percent on the predicted angular
power spectrum, so this effect should be taken into account for future
experiments like PLANCK.

\begin{figure*}
\centering
\includegraphics[angle=90,width=2.3\columnwidth,height=0.45\textheight]{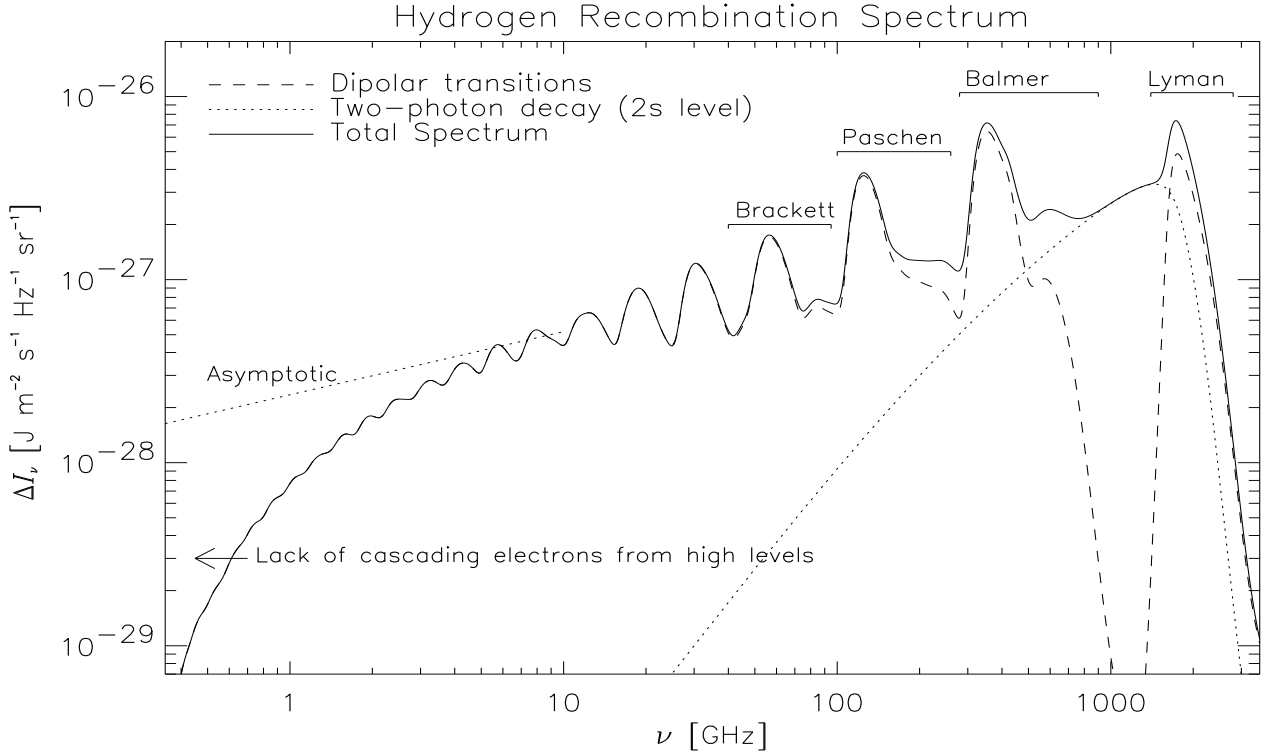}
\caption{Full hydrogen recombination spectrum for $\nmax=30$.  The
populations of all the angular momentum substates were taken into
account.  We indicated the contributions corresponding to the Lyman,
Balmer, Paschen and Brackett series. The straight line shows one
possible asymptotic behaviour at low frequencies when including higher
hydrogen shells (see text for details).}
\label{fig:main}
\end{figure*}

\begin{figure*}
\centering 
\includegraphics[width=2\columnwidth]{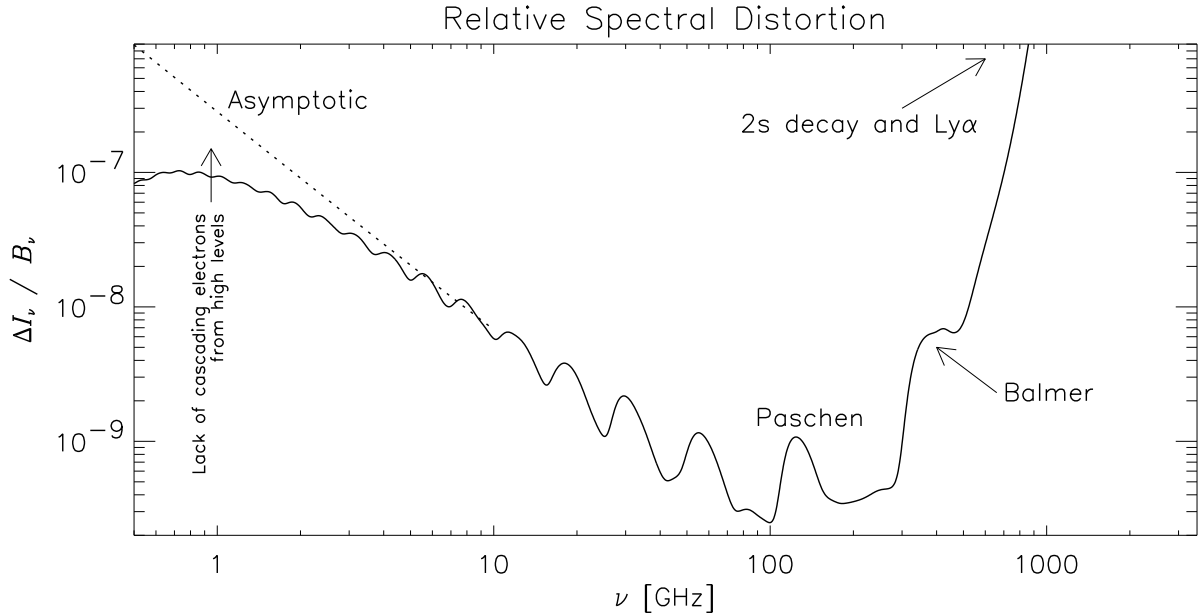}
\caption{Relative spectral distortion of the CMB spectrum for the hydrogen
recombination spectrum as computed in this paper.  We use the same frequency
range as in Fig.~\ref{fig:main}. Note that the Ly$\alpha$ and two-photon decay
features produce very strong relative distortions which exceed unity, so they
are not presented here for clarity. }
\label{fig:relative_distortion}
\end{figure*}

The main results of this paper are summarised in Fig.~\ref{fig:main} and
Fig.~\ref{fig:relative_distortion}. The
hydrogen lines from the epoch of cosmological recombination appear as 
broad ($\Delta \nu/\nu \approx 0.30-0.40$) features, 
having high contrast for the Lyman, Balmer, Paschen, Brackett and 
Pfund series.
Altogether they produce a spectral distortion with 
a characteristic oscillatory behavior, which can not be mimiced by any
other foreground, and thus might be used in order to devise
a method for their detection.
Although most of the current observational efforts are dedicated to
the study of the (temperature and polarization) angular fluctuations,
it is clear that measuring these spectral features would constitute an
unique way to test our understanding of the recombination process. For
example, it will provide a {\it direct} determination of the redshift
of recombination, and independent determinations of cosmological
parameters such as the baryon density.
%
One should mention that the broadening due to the scattering off free
electrons in the epoch of recombination is 
at most of the order of a few percent
\citep[e.g.][]{1997AstL...23..565D}, implying that the features we are
discussing in this paper are not going to be wiped out.

Cosmological recombination cannot produce many more than one photon
per transition in the hydrogen atom per one act of recombination.
This fact, together with the low entropy of the Universe, are
the reasons why only in two extreme cases might the spectral
distortions connected with the cosmological hydrogen recombination
become observable in the future: (i) the Wien part of the CMB spectrum
(Lyman-$\alpha$, 2s two-photon decay emission \& the Balmer lines) and
(ii) the Rayleigh-Jeans part.

Concerning the first case, the CIB appears six orders of magnitude
above the signals we are discussing, so the detection in the Wien part of the
spectrum will be challenging.  On the other hand, at very low frequencies the
foregrounds like the 21~cm line, synchrotron emission of relativistic
electrons of radio and ordinary galaxies become important and make
observations more complicated. From this point of view the spectral
band $\nu \ga 1.4$GHz is more interesting, because the 21~cm emission
will not contaminate in this frequency range, and 
the relative amplitude of the
distortions in this band becomes larger (of the order of
$\sim 10^{-7}$).

The paper is organised as follows. In Sec.~2 we briefly review the
asumptions of the computations of this paper, as well as the basic
equations. Sec.~3 presents the results for the complete spectrum and
for individual lines. In that section we discuss the robustness of the
results when changing the total number of shells, as well as the
non-equilibrium effects arising from the angular momentum
states. Sec.~4 and 5 present the discussion and the conclusions,
respectively.
%

\section{Basic equations}
In order to calculate the time evolution of the populations of the multi-level
hydrogen atom we adopt the procedure described in \cite{Seager2000ApJS}.
Here we would like to emphasize the most important assumptions
that this approach implies:

\begin{enumerate}
\item The observed deviations of the CMB spectrum from a blackbody are
    smaller than $\Delta I/I_0\lesssim 10^{-4}$
    \citep{1994ApJ...420..439M,1996ApJ...473..576F}. Therefore the radiation
    field is assumed to be described by a perfect blackbody with temperature
    $T_0=2.725\,$K.
\item The quasi-static solution for the spectral-line profiles is
  valid \citep{RybickiDell94}.
\item The Sobolev escape probability method \citep{sobolev} is adopted to deal
  with the evolution of all lines (see \citet{Dubrovich2004} for a derivation of
  this method in a cosmological context). In practice, this will only
  affect those transitions which are directly connecting with the ground state,
  while for the rest the optical depths $\tau_\nu$ are so small that the escape probability
  is very close to unity. Therefore it is possible to work in the optically 
  thin limit taking into account only terms proportional to $\tau_\nu$ for the corresponding transition
 \citep{Jose2005}.
\end{enumerate}

\noindent
In our computation, the major differences with respect to the
  \cite{Seager2000ApJS} procedure are:

\begin{enumerate}
\item[(i)] We split the hydrogen levels up according to their {\it principal}
and {\it angular momentum} quantum numbers $(n, l)$ and include {\it all}
sub-levels. To illustrate the dependence of the recombination spectrum 
on the treatment of the angular momentum substates, we
also consider some cases, in which we follow the populations of {\it all} 
sub-levels with $n\leq \nsplit$, while for levels with $\nsplit<n\leq\nmax$ we average over the
angular momentum states (i.e we assume that the sub-levels are populated
according to their statistical weights, so that
$N_{nl}=(2l+1)\,N_{n\rm s}$ is fulfilled). This implies that for
given $\nsplit$ and $\nmax$ we follow the populations of
$N=\nsplit[\nsplit+1]/2+\nmax-\nsplit$ separate hydrogen states.
\item[(ii)] We neglect collisional rates, as they have been found to be of
  minor importance \citep{Seager2000ApJS}. 
\item[(iii)] We only follow the HeII ionization fraction without any detailed
  follow-up of separate levels.
\item[(iv)] Instead of directly tracking the physical population $N_i$
  of a level $i$ we define the variables $X_i=N_i/N_{\rm H}$,
  where $N_{\rm H}(z)$ is the total number density of hydrogen nuclei
  in the universe. In this way one can absorb the expansion term.

\end{enumerate}

A very similar treatment has been used in \citet{Jose2005} to calculate the
imprints of the cosmological hydrogen recombination lines on the CMB angular
power spectrum including all sub-levels up to $\nmax=10$.
Here we developed two independent codes using different routines of the {\sc
  Nag}-Library\footnote{See http://www.nag.co.uk/numeric/} to solve the
coupled-system of stiff ordinary differential equations.  The results of both
codes were absolutely consistent and robust.

In addition, we explored three different numerical methods to compute the
photoionization and photo-recombination rates: the approximations (valid for
small values of the energy of the ejected electron) given by
\citet{Burgess1958}; the polynomial expressions given by \citet{Karzas1961}
and \citet{Boardman1964}; and the numerical subroutines provided by
\citet{StoreyHummer1991}. The second and third approaches were found to
give identical results for all levels up to $\nmax=30$.
Even though the approximations adopted by \citet{Burgess1958} are not valid for
high values of $n$ and frequencies far away from the photoionization threshold
of each transition, we found that for the calculations of this paper ($\nmax \le
30$), they yield similar results (within a few percent accuracy) with respect to
those obtained in the other two cases.  It also turned out that the inclusion of
induced photorecombination is important for the brightness of those
transitions with $\Delta n=1$ at high levels ($n\gg 1$).
Finally, we mention that we have also explored simple
formulae for the computation of the photoionization/photorecombination
cross-sections, as those obtained using the Kramers approximation, in
which the Gaunt factors are neglected.  We found that this kind of
approximation is {\it not} reproducing properly the trends within a
given shell, overestimating the rates which are connecting those
sublevels with very high $l$-values to the continuum. 

For all the computations presented in this work the following values of
  the cosmological parameters were adopted \citep{WMAP_params}:
  $\Omega_{\rm b} =
  0.0444$, $\Omega_{\rm tot}=1$, $\Omega_{\rm m}=0.2678$, $\Omega_{\Lambda}=0.7322$,
  $Y_{\rm p}=0.24$ and $h=0.71$.

%
\subsection{Hydrogen spectral distortions due to radiative dipole transitions}
In order to calculate the spectral distortions arising from the
hydrogen dipole transitions we use a $\delta$-function approximation
for the line-profile, i.e. we neglect any lifetime effects
(Lorentz-profile) or thermal broadening (Voigt-profile).
A straightforward derivation yields the spectral distortion, $\Delta
I_{ij}(\nu)$, arising form the transition of the upper level $i$ to the lower
level $j$ \citep[e.g. also see][]{Wong2005}
\beal
\label{eq:I_ij}
\Delta I_{ij}(\nu)=\frac{ch}{4\pi}\frac{\Delta R_{ij}(\zem)}{H(\zem)[1+\zem]^3}
\Abst{,}
\end{align}
where $\zem$ is the redshift of emission, which relates the observing
frequency $\nu$ to the transition frequency $\nu_{ij}$ by
$\nu=\nu_{ij}/[1+\zem]$. Furthermore $H$ denotes the Hubble-factor and the
effective radiative rate for the considered transition is given by
\beal
\label{eq:DRij}
\Delta R_{ij}=p_{ij}\,\frac{A_{ij}\,N_i\,e^{h\nu_{ij}/\kB\Tg}}{e^{h\nu_{ij}/\kB\Tg}-1}
\left[1-\frac{g_i}{g_j}\,\frac{N_j}{N_i}\,e^{-h\nu_{ij}/\kB\Tg}\right]
\Abst{,}
\end{align}
where $p_{ij}$ is the Sobolev-escape probability as defined in
\citet{Seager2000ApJS}, $A_{ij}$ is the Einstein-$A$-coefficient of the
transition, $N_i$ and $g_i$ are the population and statistical weight of the
upper and $N_j$ and $g_j$ of the lower hydrogen level, respectively.
Furthermore we have assumed that the ambient photon field is given by a blackbody spectrum with temperature $\Tg=T_0[1+z]$, where $T_0=2.725\,$K is the
present CMB temperature \citep{Fixsen2002}, and we have made use of the
Einstein-relations.
%

%
\subsection{Hydrogen 2s-1s spectral distortion}

The spectral distortion arising due to the two-photon decay of the hydrogen 2s
level can be obtained by the integral \citep[e.g. see][]{ZeldKurtSun1968orig, Wong2005}
\beal
\label{eq:I_2s1s}
\Delta I_{\rm 2s1s}(\nu)=\frac{ch\nu}{4\pi}\int_0^\infty 
\frac{\Delta R_{\rm 2s1s}(z)\,\phi(\nu[1+z]/\nu_{\alpha})}{H(z)[1+z]^3}\id z
\Abst{.}
\end{align}
Here $\Delta R_{\rm 2s1s}=A_{\rm 2s1s}\,[N_{2s}-N_{\rm 1s}e^{-h\nu_{\alpha}/k
  T_{\gamma}}]$, where the total two-photon decay rate for the hydrogen
2s-level is given by $A_{\rm 2s1s}=8.22458\,\text{s}^{-1}$ and
$\nu_{\alpha}\approx \pot{2.46}{15}\,$Hz is the Lyman-$\alpha$ photon
frequency.

A definition of the profile function $\phi(y)$ can be found in
\citet{Spitzer1951} but a sufficient approximation was obtained by
\citet{Nussbaumer1984}
\beal
\label{eq:phi_appr}
\phi(y)=\frac{C}{\nu_{\alpha}}\,\left[w(1-4^\gamma\,w^\gamma)+\alpha\,w^{\beta+\gamma}\,4^\gamma\right]
\Abst{,}
\end{align}
where $w=y[1-y]$, $C=24.5561$, $\alpha=0.88$, $\beta=1.53$ and $\gamma=0.8$.
Here it is important to note that $\phi(y)$ is normalized to
$\int_0^{\nu_{\alpha}} \phi(\nu/\nu_{\alpha})\id \nu =2$ in order to account
for the fact that {\it two} photons are emitted per decay of {\it one}
hydrogen 2s-level \citep[see also][]{Boschan1998}.

\section{Results}

\subsection{The full recombination spectrum}
In Figures~\ref{fig:main} and \ref{fig:relative_distortion} we present
the full hydrogen recombination spectrum for $\nmax=30$. The
population dependence on $l$ was taken into account for all shells
($\nsplit=30$).  In this calculation the evolution of $465$ separate
hydrogen states in the redshift range of $500\leq z\leq 3500$ was
included.

At high frequencies one can clearly see the features corresponding to
the Lyman, Balmer, Paschen and Brackett series. We will discuss some
of these lines in more detail below. At lower frequencies the lines
start to overlap strongly and eventually merge to a continuum at very
low $\nu$.
The figure also shows one {\it possible} asymptotic behavior at low
frequencies.  This is meant to illustrate the fact that, when
including a larger number of shells, the additional cascading
electrons will enhance the emission at low frequencies.  Here we
decided to plot this asymptotic limit using the slope of
\citet{Kholu2005}. These authors were the first to obtain the spectral
distortions for a very high number of levels ($\nmax=160$), although
under the assumption of quasistatic evolution and without taking into
account the populations of the angular momentum substates seperately.
Their average slope in the 1~GHz to 10~GHz range is approximately
$+0.35$, while in our case, the average slope in the frequency range
$2-10\,$GHz has a value of approximately $+0.8$.
A detailed discussion of this asymptotic region requires the inclusion
of several physical effects which have not been treated in this
paper. For example, it is clear that at even lower frequencies
free-free absorption will start to erase any emission from the very
high hydrogen levels, and in addition collisions should become
important.  Therefore, the study of the low-frequency asymptotic
region will be left for a subsequent paper. However, we expect that
the value of the slope obtained from the paper of \citet{Kholu2005}
provides a lower limit.


Table~\ref{tab:lines} summarises some parameters (peak amplitude,
central frequency and relative width) of the first five features in
the complete (including the 2s decay line) spectrum presented in
Fig.~\ref{fig:main}.  It is interesting to compare these values with
those for the corresponding $\alpha$-transition ($\Delta
n=1$). Clearly, the main contribution to the features comes from this
line, although the emission due to the 2s-decay is significant in the
range of the Lyman-series.  Furthermore, the width of the features is
slightly increased with respect to the $\alpha$-line due to the
addition of higher transitions ($\Delta n>1$), and in the case of the
Lyman and Balmer features, because of the 2s decay emission.

\begin{table*}
\caption{Positions and amplitudes of the main features shown 
in Fig.~\ref{fig:main}.
We show the central frequency ($\nu_0$) and wavelength ($\lambda_0$) 
as observed today, the relative 
width at half maximum ($\Delta \nu/\nu$) and the peak 
intensity for the features corresponding to the first five series
of hydrogen. 
For comparison, the last five columns show the central 
redshift of formation, the central frequency, the relative width, the peak
and relative amplitude of the corresponding
$\alpha$-transition (i.e. $\Delta n=1$) when considered separately. }
\label{tab:lines}
\centering
\begin{tabular}{@{}lcccccccccc}
\hline
\hline
Series & $n$ & $\nu_{0}$ &  $\lambda_0$ & $\Delta \nu / \nu$  & $\Delta I_\nu$(peak) &  
$z_{\alpha}$ & $\nu_{\alpha}$ &  ($\Delta \nu / \nu$)$_\alpha$ 
& $\Delta I_{\alpha}$ & $\Delta I_{\alpha}/ \Delta I_\nu$ \\
       &     &  [GHz]   & [$\mu$m] &      &    
$[{\rm J\,m^{-2}\,s^{-1}\,Hz^{-1}\,sr^{-1}}]$ &  & [GHz]  &  & 
$[{\rm J\,m^{-2}\,s^{-1}\,Hz^{-1}\,sr^{-1}}]$ & [\%] \\
\hline
Lyman & 1  &   1732 & 173  &  
0.28 &    $\pot{7.4}{-27}$ &      1407  &  1752 &   0.23 & $\pot{4.8}{-27}$ & 65\\
Balmer & 2  &       353  & 849  
 &  0.37 &    $\pot{7.2}{-27}$ &  1274 &  358  &   0.26 & $\pot{6.5}{-27}$ & 93 \\
Paschen & 3  &       125  &  2398 & 0.31  &    $\pot{3.8}{-27}$ &  1259  &  
127 &   0.26 & $\pot{3.3}{-27}$ & 87 \\
Brackett & 4  &       56  &  5353 & 0.36 &    $\pot{1.8}{-27}$ &  1304 &   
  57  &  0.27 &  $\pot{1.2}{-27}$ & 67 \\
Pfund & 5  &       30  &  9993 & 0.39 &    $\pot{1.2}{-27}$ &  1312 &  
31 &   0.27 & $\pot{7.3}{-28}$ & 61 \\
\hline
\hline
\end{tabular}
\end{table*}

\begin{figure}
\centering 
\includegraphics[width=\columnwidth]{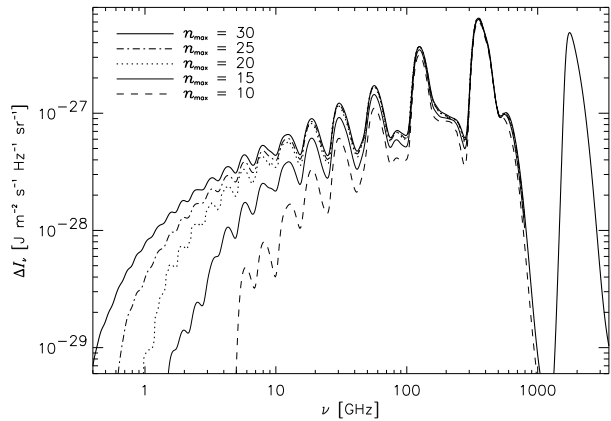}
\caption{Hydrogen recombination spectrum for different values of $\nmax$. The
  contribution from the 2s two-photon decay was not shown on this
  plot. We show the cases of $\nmax = 10, 15, 20, 25$ and $30$, always with
  $\nsplit=\nmax$.}
\label{fig:convergence}
\end{figure}

\subsubsection{Dependence on the number of shells}
At frequencies $\nu\lesssim 2$~GHz the slope of the computed spectrum changes
significantly. This is due to the lack of diffusion of electrons from levels
above $\nmax$ and transitions from those levels with $\Delta n>1$, 
which have not been accounted for here.
In order to estimate this effect we have performed additional
calculations with $\nmax=10,\, 15,\,20$ and $25$, all for
$\nsplit=\nmax$.  A compilation of the results is shown in
Figure~\ref{fig:convergence}.  One can see that the lines of the
Lyman, Balmer, Paschen and Brackett series basically do not
change when increasing the number of shells beyond $\nmax\sim 20$.
On the other hand, the low frequencies part ($\nu\lesssim10\,$GHz) is
still varying rather significantly. Our calculations suggest that for
$\nmax=30$ at the level of a few percent the spectrum is
converged for $\nu\gtrsim 20\,$GHz even when adding more
shells. Pushing $\nmax$ to larger values should lead to an increase of
the distortions at low frequencies and eventually should partially
fill the gap with respect to the asymptotic behaviour
indicated in Figure~\ref{fig:main}.

\begin{figure}
\centering 
\includegraphics[width=\columnwidth]{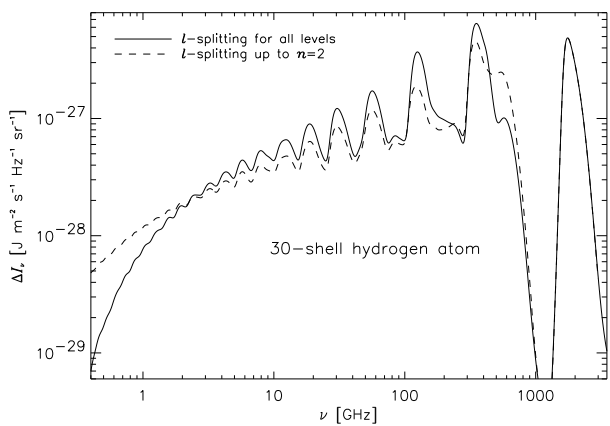}
\caption{Non-equilibrium effects of the angular momentum states on the 
  hydrogen recombination spectrum. We consider the case $\nmax=30$ 
  for two different values of
  $\nsplit$ as indicated in the legend. The contribution from the 2s
  two-photon decay was omitted here.}
\label{fig:splitting}
\end{figure}

\begin{figure*}
\centering 
\includegraphics[width=\columnwidth]{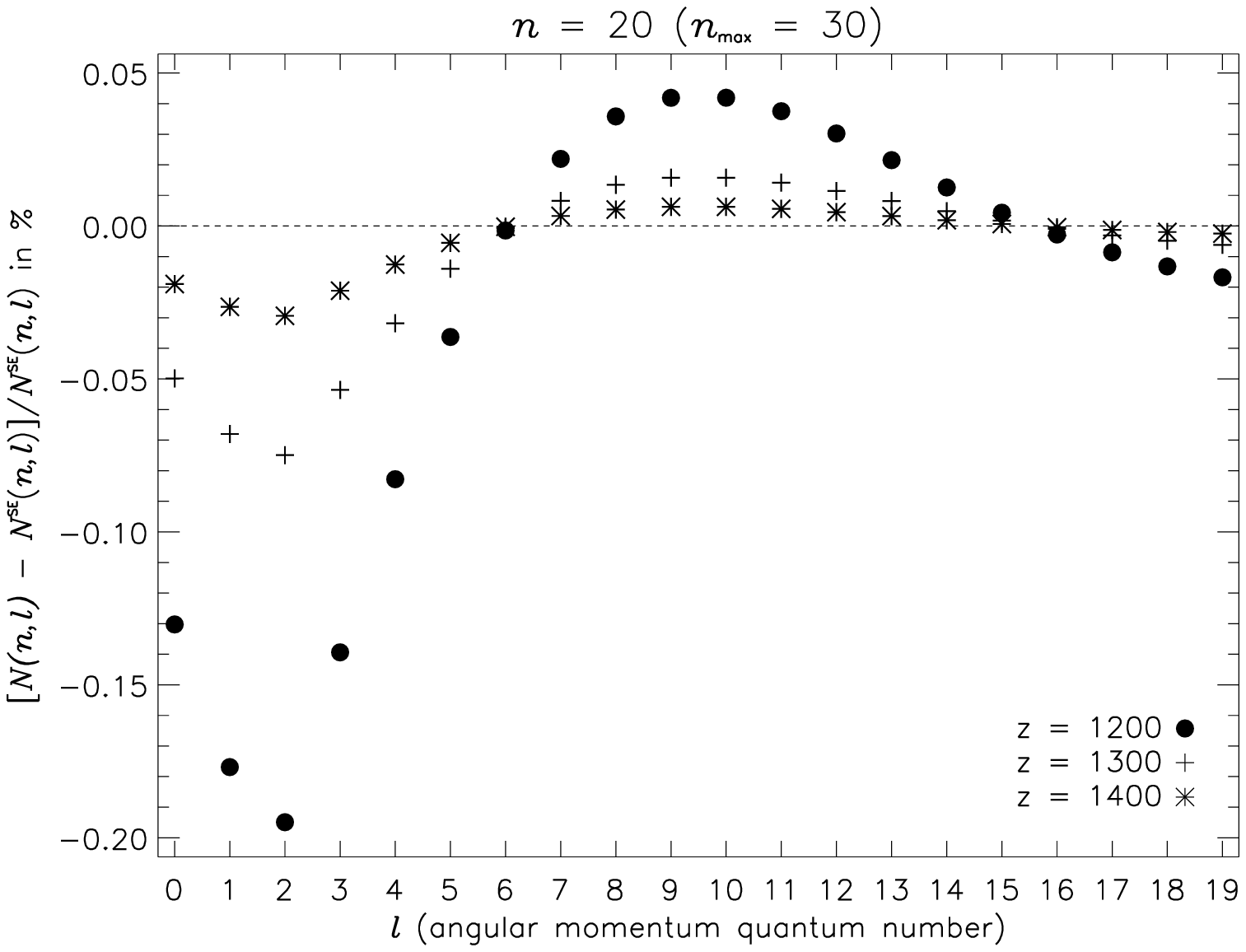}%
\includegraphics[width=\columnwidth]{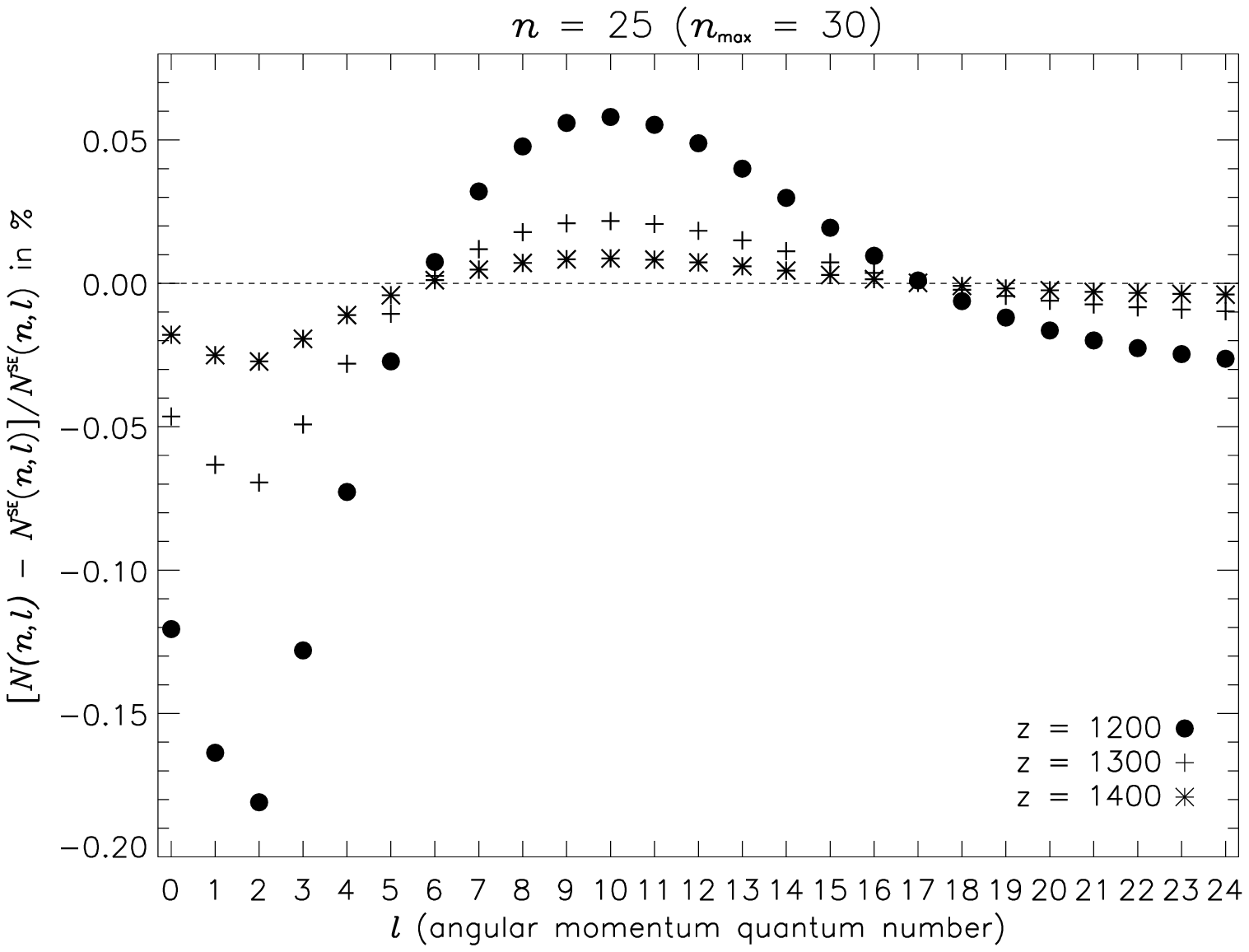}
\includegraphics[width=\columnwidth]{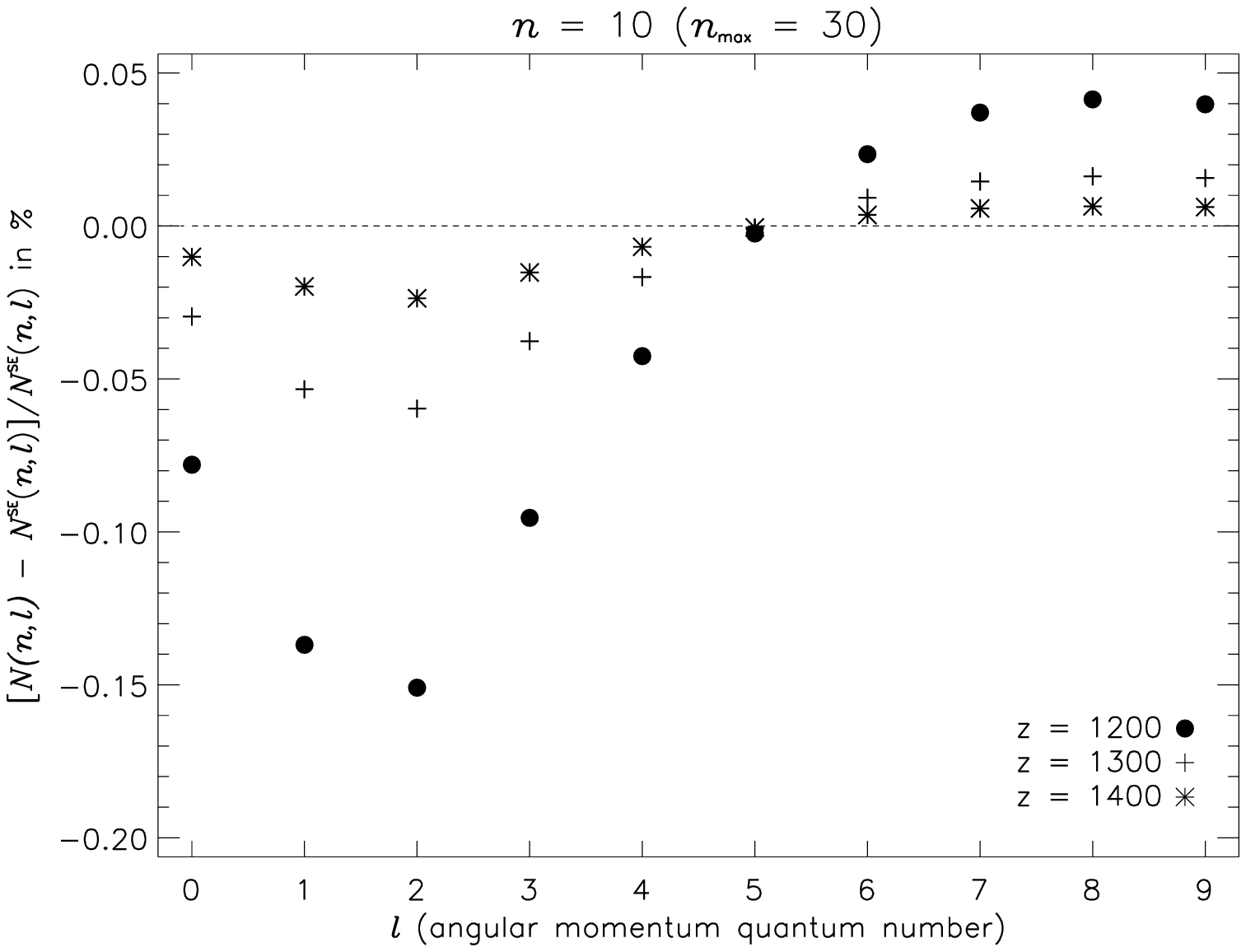}%
\includegraphics[width=\columnwidth]{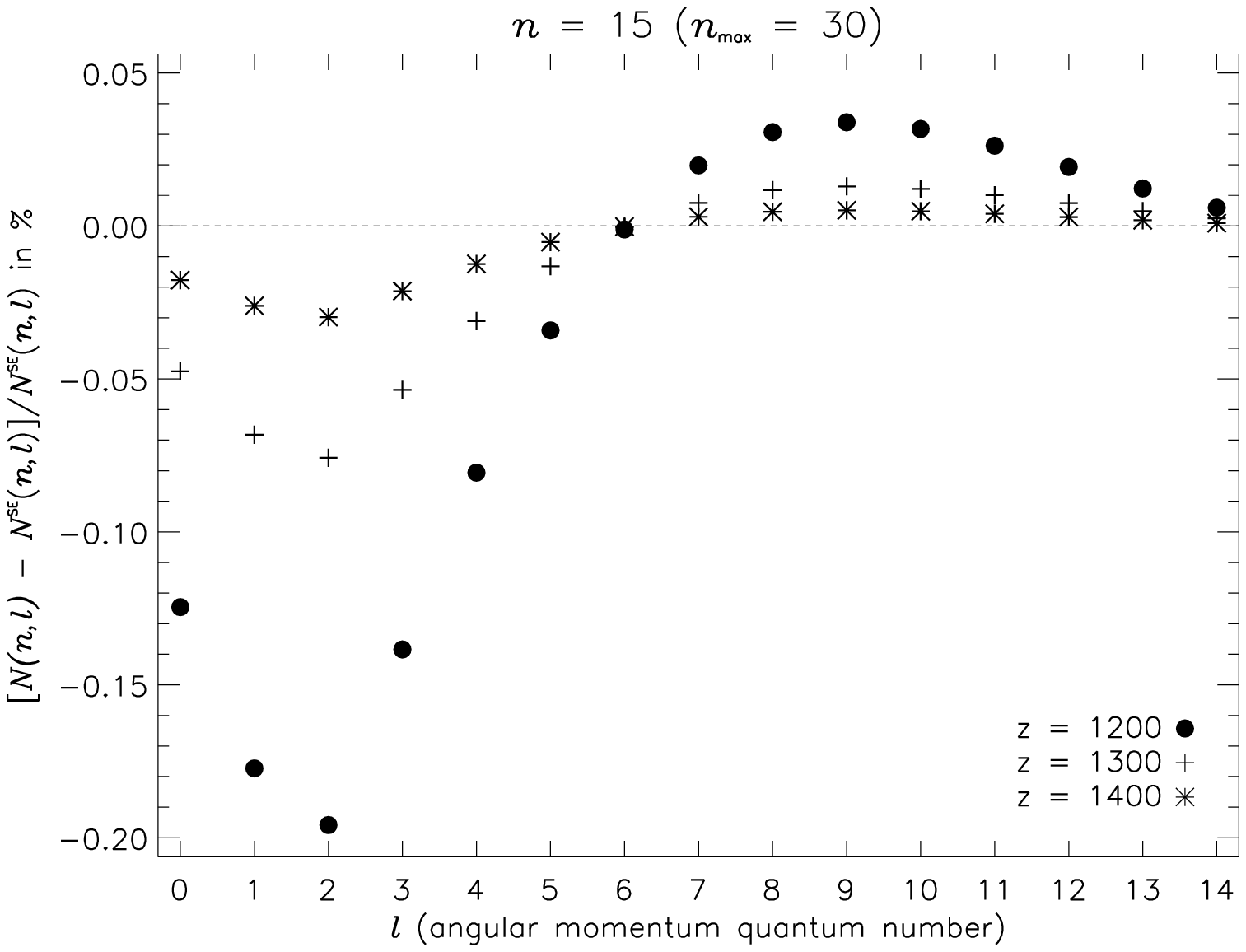}
\includegraphics[width=\columnwidth]{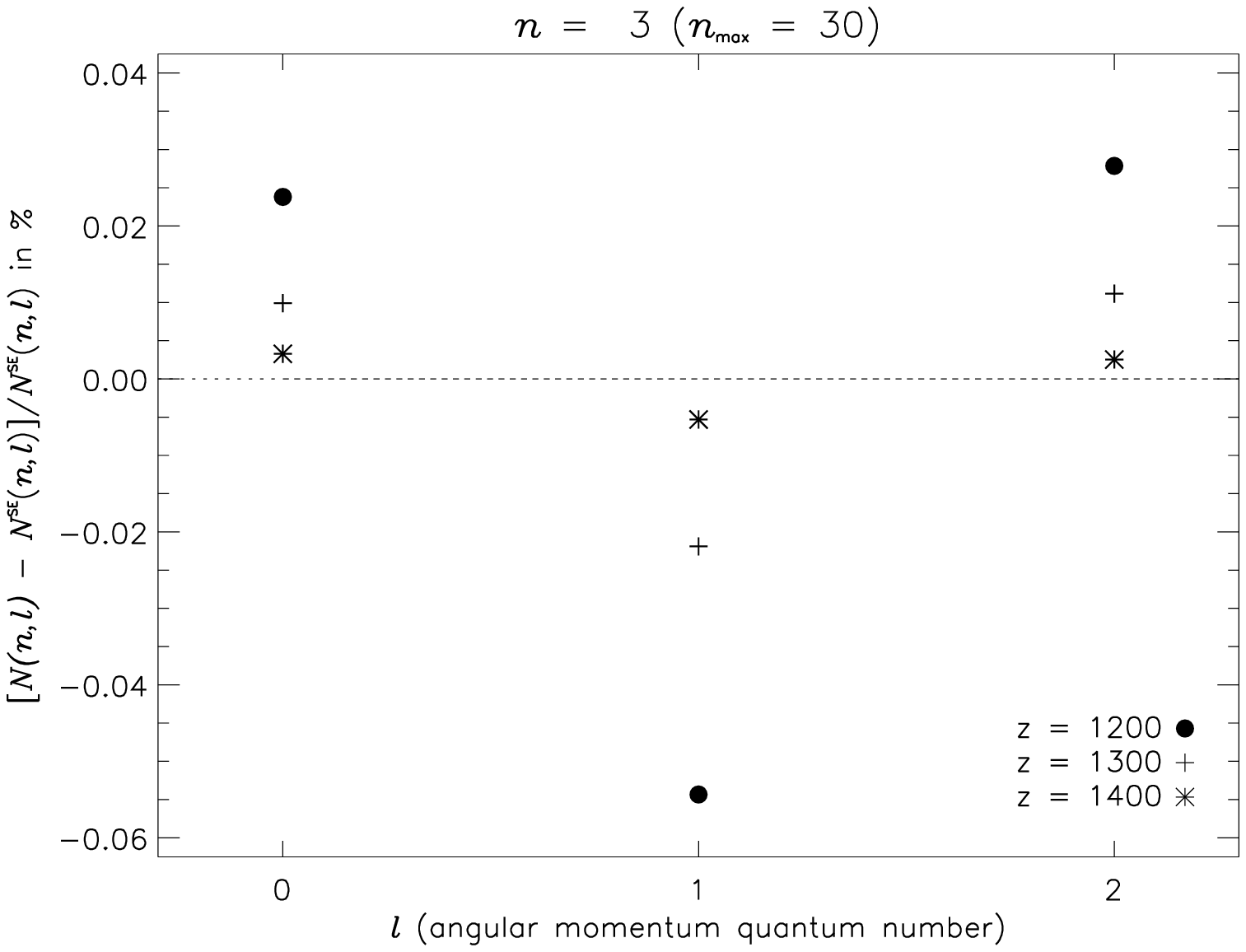}%
\includegraphics[width=\columnwidth]{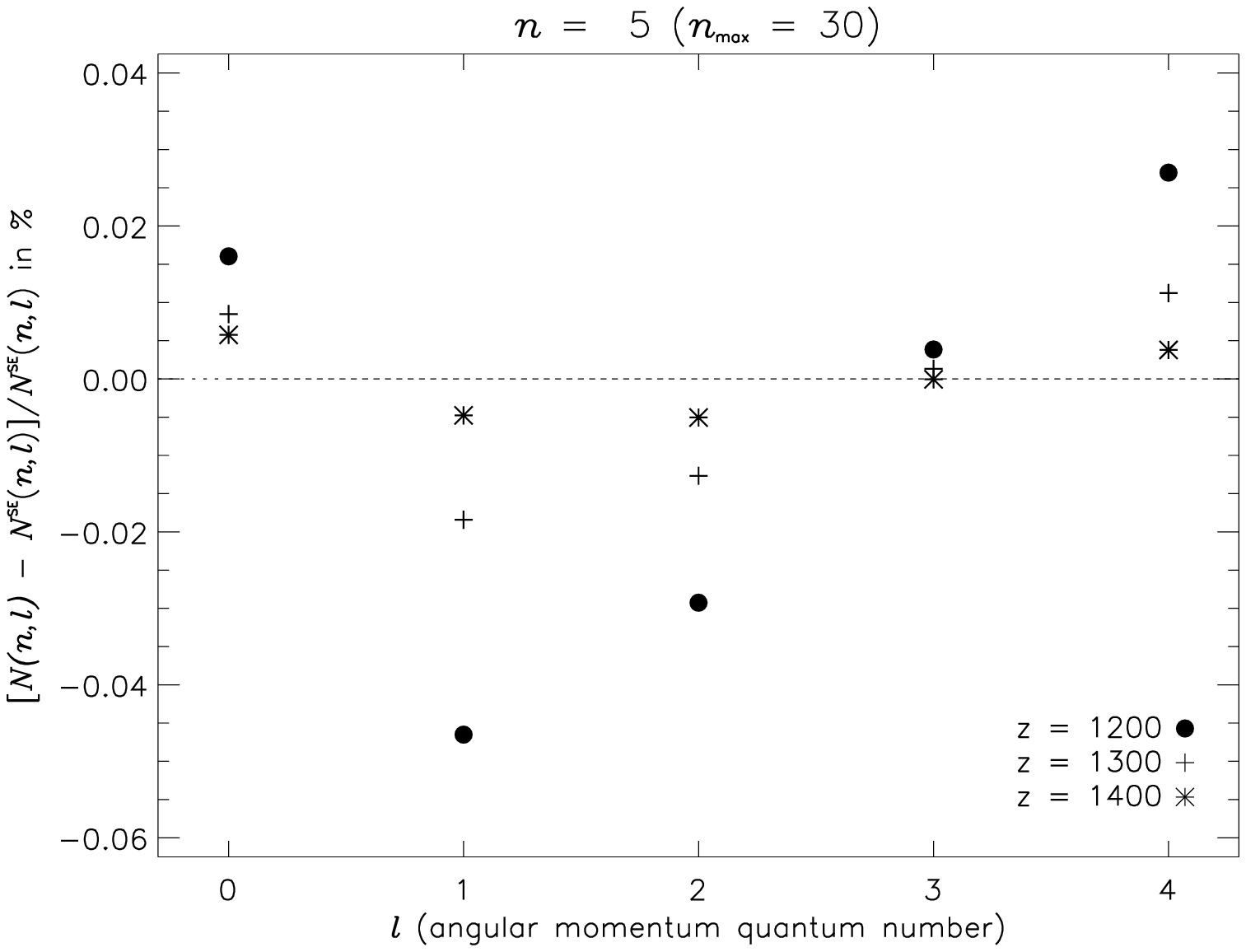}
\caption{Non-equilibrium effects on the populations of the angular
momentum substates for a given $n$. The results were obtained for
$\nmax=30$ following the populations of all the angular momentum
sub-states.
    We present the ratio $N_{n l} / N^{\rm SE}_{n l}$, where the
    statistical equilibrium (SE) population is computed from the
    actual total population of the shell by $N^{\rm SE}_{n l} =
    [(2l+1)/n^2] N_{\rm tot}$. Six cases are shown, corresponding to
    values of the principal quantum number $n=$ 3, 5, 10, 15, 20 and
    25.}
\label{fig:nLTE_to_n}
\end{figure*}

\subsubsection{Non-equilibrium effects of the angular momentum states}
Most of the computations of the hydrogen recombination spectrum
published to date were either based on the effective three-level atom
\citep[e.g.][]{Boschan1998, Wong2005}, or used averaging over angular
momentum sub-levels \citep[e.g.][]{Dubrovich2004, Kholu2005}.  Here we
shall discuss the influence of the latter simplification on the
solution obtained for the recombination spectrum.

Fig.~\ref{fig:splitting} compares our result for $\nsplit=30$ with the
case of $\nsplit=2$ (which represents the standard computation without
detailed follow-up of the $l$-substates except for the 2s and 2p
levels\footnote{Hereafter, we will refer to the case $\nsplit=2$ as
the ``non-splitting'' solution, and to the case $\nsplit=\nmax$ as the
full follow-up case.}).
The ``non-splitting'' solution has a higher intensity at frequencies
smaller than $\nu \sim 2$~GHz and also a smaller slope in this
region. This shows that, for a hydrogen atom with given $\nmax$, the
$\nsplit=2$ solution is effectively having a larger number of
transitions in the upper shells (i.e. higher diffusion), resulting in
an excess of emission.  We will come back to this point below.
In addition, there are also effects on the high frequency
part of the spectrum. For $\nsplit=2$ the peak intensities of the main
transitions ($\Delta n = 1$) are under-estimated while the ratios of
the main to the secondary transitions ($\Delta n>1$) are decreased. In
other words, the incomplete treatment of the high levels also
affects the diffusion to lower levels, and hence the line intensities
are affected as well.
For example, the Balmer lines of the spectrum are strongly
affected. The ratio of the two peaks appearing at $\sim 350$~GHz and
$\sim 550$~GHz (which correspond to H$\alpha$ and a superposition of
lines higher than H$\gamma$, respectively; see the Section on Balmer
lines) decreases from roughly 6 to about 2, and the height of the
smaller peak is more than 2 times bigger than in the full calculation.

We note that a comparison of our results for $\nsplit=2$ with those
obtained by \citet{Kholu2005} shows a very good correspondence.  When
doing the computations using their cosmological parameters, we find
good agreement also in the amplitudes, showing that our two
implementations can produce similar results.  
Given that these authors produced their spectrum assuming a
quasistatic-evolution for all shells above $n=2$, our results further
supports the idea that their approach is appropriate
\citep[see also the comments in][]{2004AstL...30..509D}.
However, including the evolution of the populations of all
$l$-substates in this approach seems more difficult to solve, because
the deviations from statistical equilibrium within a shell are very
small, and the corresponding columns in the matrix of the level
populations will be degenerate.
To test this point, we have also implemented the quasi-static approach
for shells $n>2$, while solving the evolution (differential equations)
for $n\le 2$ and the electron fraction.
For our implementation of the problem, we found that, 
when considering all the angular momentum substates separately,
with standard methods solving the complete set of ordinary differential 
equations seems more efficient (in terms of computational time) than 
solving the corresponding algebraic system.

Finally, we have also performed several computations considering values of
$\nsplit$ in the range between 2 and $\nmax$, mainly with the aim of
understanding whether the full computations could be simplified.  
In these cases, we found that the spectrum obtained is closer
to the full solution ($\nsplit=\nmax$) in the high frequency part.
However, there always appears an abrupt change of slope in the
spectrum precisely at the frequency corresponding to the main
transition of the $\nsplit$-series. Below this frequency, the average
slope of the spectrum is flatter than in the full solution.

Obviously, all the effects discussed in Fig.~\ref{fig:splitting}
appear because the inclusion of $l$-substates modifies
the populations of the levels.
In order to understand these modifications, and how the differences in
the populations of the levels are connected with changes in the
spectral distortions, we shall now discuss these effects in detail.

\paragraph*{Effects within a given shell.}
We first discuss the situation for a given shell (i.e. we fix $n$),
where we compare the populations of the different sub-levels with the
corresponding expected value in the case of statistical equilibrium
(SE) within the shell, i.e. when $N_{n l}^{\rm SE}= [(2l+1)/n^2]
N_{\rm tot}$, where $N_{\rm tot}$ is the (actual) total population of
the shell.

In Fig.~\ref{fig:nLTE_to_n} we present the ratios $N_{n l} / N^{\rm
SE}_{n l}$ of the true population $N_{n l}$ of the level for
the full computation with $\nmax=30$, to the expected SE population.
We show six different cases, corresponding to values of the principal
quantum number $n=3$, $5$, $10$, $15$, $20$ and $25$. For each one of these
cases, we present the ratios at three different redshifts
($z=1200$, $1300$ and $1400$) which correspond to the region where
the lines are forming.
At higher redshifts, the deviations from statistical equilibrium
show a similar trend, but they are much smaller in amplitude
and tend to dissappear. This behaviour is expected since one
is then approaching the epoch of full thermodynamic equilibrium
\citep{ZeldKurtSun1968orig}.
In our calculations the non-equilibrium effects are compatible with zero
within the numerical precision of the two codes at $z \gtrsim 2000$.

%
The general trend for all shells above $n=5$ can be summarized as
follows:
within a given shell the sub-levels with $l \la 5-6$ are
underpopulated as compared to $N^{\rm SE}_{n l}$, with the difference
being the strongest for the d-level ($l=2$) and smallest for the level
$l\sim 5-6$.
This shows that in the calculations with full follow-up these angular
momentum states depopulate faster.
For $l > 6$ the population of the sub-states is larger than in the SE
situations\footnote{Note that this is not true for the very 
high $l$-substates and the outer shells, as $n=20$ or $n=25$, 
as a consequence of the lack of cascading electrons from higher levels.}.
This implies that these angular momentum states depopulate slower,
which is likely due to blocking of the transition to lower levels 
($\Delta l=\pm 1$ restriction).
On the other hand, for those shells $n \le 5$, the p and d-levels are
also underpopulated, but now the strongest deviation appears for
p-level ($l=1$).
When comparing our results for $\nmax=30$ with the case of $\nmax=25$
and $\nmax=20$, we can conclude that the results in
Fig.~\ref{fig:nLTE_to_n} for $n\le 10$ have already converged, but
higher shells could increase their deviations from SE,
especially at high $l$-states.

Although the explanation of these non-equilibrium effect requires
one to study the net rates for all transitions connecting to
that particular level, we can understand some of the general trends
with the following arguments.
It is clear that assuming SE is equivalent to ``instantaneous''
redistribution within a given shell, i.e. if an electron is captured
by a given $l$-state, this will immediately increase the population of
{\it all} the other $l$-states.
One would expect that instantaneous redistribution will operate when
collisions $(n,l)\rightarrow(n,l')$ are effective. This should
eventually happen in sufficiently high shells (see
Sect.~\ref{sec:discusion}).  On the other hand, when instantaneous
redistribution is not working, then the depopulation of each level
will depend on the exact route the electron can take to reach the
ground state.

As the Universe expands, the blackbody radiation field is redshifting
to lower energies, and thus the number density of high energy photons
decreases with time.  When the radiation field does not contains
enough photons to maintain equilibrium in a given shell $n$ (i.e. when
the number of photons with energy above the photoionization potential
of the shell $\chi_n$ falls below 1 per atom), then the
photoionization rates become smaller than the photorecombination rates
and the electrons can not escape to the continuum. Note that this is
first happening for lower shells with the largest photoionization
potential.

In the above situation, the electrons try to cascade down to lower
levels. Our computation only includes radiative rates, so electrons in
a given shell can only diffuse down through transitions with $\Delta l
=1$.  In general, it seems that the fastest transitions are found to
be those which can directly connect with the 2s and 2p-levels, so it
is reasonable to expect that all sub-levels with $l\le 2$ will be able
to depopulate faster in the complete ($\nsplit=\nmax$) computation,
and their SE populations will seem larger.
On the other hand, due to the bottle-neck, which is produced in the
lower states during cosmological recombination, states with high
values of $l$ depopulate slower in the full computation, since
electrons have to cascade down via many intermediate shells.

\begin{figure}
  \centering \includegraphics[width=\columnwidth]{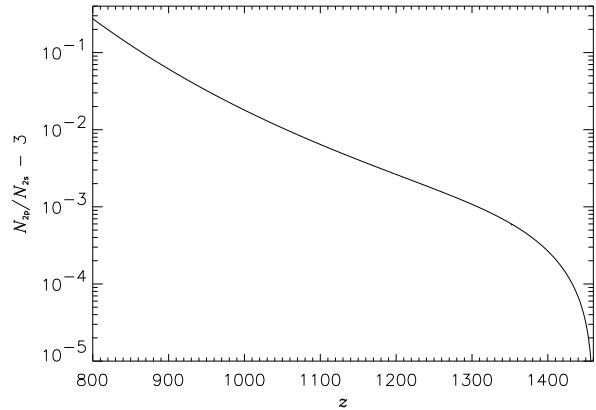}
\caption{Non-equilibrium effects in the populations of the 2s and 2p levels.  We
  present the ratio of the 2s and 2p populations referred to the equilibrium
  ratio (i.e. the statistical weights), as a function of redshift. Although the
  deviations are very small, they indicate that there is a deficit of electrons
  in the 2s state as compared to the 2p state as a consequence of the
  recombination dynamics.  }
\label{fig:2s2p}
\end{figure}

Finally, we also present in Fig.~\ref{fig:2s2p} the ratio of the populations of
the levels 2s and 2p as a function of redshift. This figure quantifies the fact
that at high redshifts, both the 2s and 2p populations are in statistical
equilibrium, so their population ratio equals the ratio of the statistical
weights. However, as the recombination proceeds, deviations from the equilibrium
appear, in the sense that there is an overpopulation of the 2p level with
respect to the 2s level because the main channel of recombination at low
redshifts ($z \la 1400$) is the 2s two photon decay. We will return to this
point in Sec~3.2 and 3.3.

\begin{figure}
  \centering \includegraphics[width=\columnwidth]{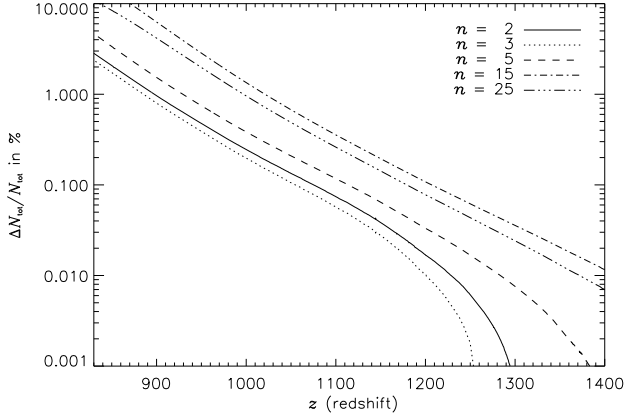}
\caption{Non-equilibrium effects in the populations of the 
  levels for different values of $n$.  We present the relative change in the
  total population of the levels ($N_{\rm tot}(n) = \sum_l N_{n l}$) due to
  the detailed follow-up of the angular momentum sub-states, for a hydrogen
  atom with $\nmax=30$.  Note that the fractional change, defined as
  $[ N^{(\nsplit=\nmax)}_{\rm tot} - N^{(\nsplit=2)}_{\rm tot} ]/
  N^{(\nsplit=2)}_{\rm tot}$, 
  is always positive, so that the total population of a given shell in the
  calculations for $\nsplit=\nmax$ is always larger than for $\nsplit = 2$. }
\label{fig:ratio_n}
\end{figure}

\paragraph*{Effects on the total population of different shells. }
The effects connected with the departure from SE within a given shell
are not only visible for different sub-states, but they also produce a
net effect on the overall population of the level.
We illustrate this point in Fig.~\ref{fig:ratio_n}, where we present the
relative change of the overall population of a given shell for $\nmax=\nsplit$, 
with respect to the the standard ($\nsplit=2$) computation.

For a given shell $n$, the total population (sum over all $l$ states) in the
computations with $\nsplit=\nmax$ is always larger than for $\nsplit =
2$. This shows that in general the overall depopulation of a given
shell is slower for full follow-up.
Moreover, for $n>3$ the relative difference becomes larger when increasing the
value of $n$ (for $n=3$ the net effect is slightly smaller than for level
$n=2$).  For high values of $n$ (close to $\nmax$), this relative change seems
to decrease again as a consequence of the lack of diffusion
from higher shells.

At low redshifts $z \la 900$ the relative difference becomes larger than 1\%
for all the levels. On the other hand, in the redshift range where
the spectral distortions are produced (e.g. for the main transitions, the
redshift of formation is $z\sim 1300$ for high $n$, and $z\sim 1400$ for
$n=2$), the fractional change is extremely small.  This shows that the total
population of the shell during the epoch of recombination does not depend
significantly on the treatment of the angular momentum sub-states, although
these small deviations are sufficient to produce the effects shown in
Fig.~\ref{fig:splitting}.

%
\paragraph*{Effects on the transitions between different shells }
In this paragraph we wish to understand the effects on the transition 
rates between different shells.
Although the detailed analysis again would require one to show the
differences in the net rate for each transition, we will show that
using the results and arguments of the last two paragraphs we can
understand the general trend and sign of the correction to the
resulting spectral distortion as shown in Fig.~\ref{fig:splitting}.

When including $l$-substates, the populations of all
levels will be given by $N_i \equiv N_i^{\rm ns} + \Delta N_i$, where
we introduced $N_i^{\rm ns}$ referring to the populations in the case
of `non-splitting' in angular momentum above $n=2$, i.e. $\nsplit=2$.
Given that the optical depth for all transitions to levels above the
ground state is very small \citep{Jose2005}, the escape probabilities
$p_{ij}$ are very close to unity and therefore will practically not
change when following the populations of the $l$-substates.
Thus, the net rate can be written as $\Delta R_{ij}=\Delta R_{ij}^{\rm
ns}\,[1+\Delta]$, where $\Delta R_{ij}^{\rm ns}$ is the rate as
obtained in the calculation for $\nsplit=2$. The function $\Delta$ is
defined by
\begin{equation}
\Delta R_{ij}^{\rm ns}\times\Delta=\frac{\Delta N_j}{N_j^{\rm ns}}
\left[ \frac{(\Delta N_i/N_i^{\rm ns})}{(\Delta N_j/N_j^{\rm ns})}
-\frac{g_i}{g_j}\,\frac{N_j^{\rm ns}}{N_i^{\rm
ns}}\,e^{-h\nu_{ij}/\kB\Tg}\right] \Abst{.}
\label{eq:corr}
\end{equation}
For simplicity, let us now only consider the
$\alpha$-transitions. These give the main contribution to the CMB
spectral distortion and always lead to {\it positive} features (see
Fig.\ref{fig:splitting}), implying that
\begin{equation}
\Delta R_{ij}^{\rm ns}\propto
\left[ 1
-\frac{g_i}{g_j}\,\frac{N_j^{ns}}{N_i^{ns}}\,e^{-h\nu_{ij}/\kB\Tg}\right] > 0
\Abst{.}
\label{eq:corr_ns}
\end{equation}
Therefore the sign of $\Delta$ will be defined by the rhs of
Eq. \eqref{eq:corr}.

Now, comparing Eq.~\eqref{eq:corr} with \eqref{eq:corr_ns} one can
conclude that for $(\Delta N_i/N_i^{\rm ns})/(\Delta N_j/N_j^{\rm ns})
> 1$ the term within the brackets is positive, and becomes negative if
this ratio is smaller than $1$. In addition one has to check the sign
of the factor $\Delta N_j/N_j^{\rm ns}$ to obtain the overall sign of
$\Delta$.

A detailed analysis would require one to study the sign and
amplitude of these corrections for every one of the different channels
contributing to a given transition. These numbers can in principle be
derived by combining the information of
Fig.~\ref{fig:nLTE_to_n} and \ref{fig:ratio_n}.  However, it is clear
that a better approach would be to directly study the net rates for
every transition of interest.  We will do that in the following
subsections for some examples, although here we want to illustrate
that one can understand the overall sign of the correction to the
spectral distortions with this argument.

If we apply Eq.~\ref{eq:corr} to the {\it total} populations of the
two levels that we are connecting (this means that we neglect the
deviations from SE within the shell as have been discussed above),
then we can use Fig.~\ref{fig:ratio_n} to understand the net sign of
the correction. All the fractional changes in Fig.~\ref{fig:ratio_n}
are positive, so we only need to consider the ratio between the levels
involved in the transition.
On one hand, for $2< n \la 20$ the relative population is growing
with $n$, so Eq.~\ref{eq:corr} shows that the correction term to the
net rate should be positive, and thus the spectral distortion in the
full computation lies above the result for non-splitting.  However, for
high $n$-values (close to $\nmax$), Fig.~\ref{fig:ratio_n} shows that
$(\Delta N_i/N_i^{\rm ns})/(\Delta N_j/N_j^{\rm ns})$ can be expected
to be less than unity for $n\ga 20$, implying that the term in
Eq.~\ref{eq:corr} inside the brackets can eventually become negative,
and thus the spectral distortion in the full computation will lie
below the non-splitting solution at low frequencies.
%

We note that the above argument is only used here to show the
qualitative behavior of the correction when taking into account
angular momentum sub-states.  In our real computations, we considered
all possible radiative channels (and hence all possible sub-levels),
which contribute to a given transition, so a detailed analysis based
on the net rates $\Delta R_{ij}$ for all channels would in principle
be possible.

\begin{figure}
\centering 
\includegraphics[width=\columnwidth]{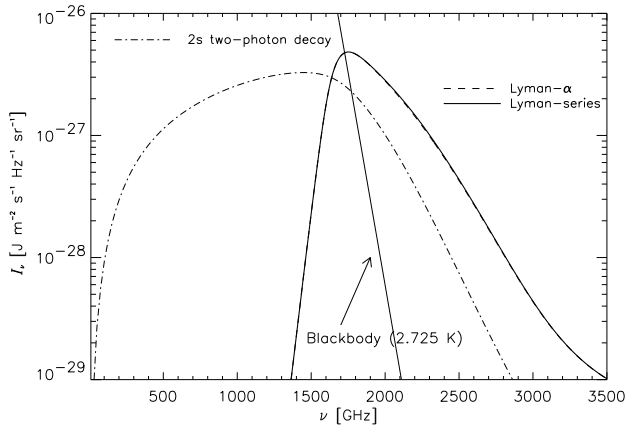}
\caption{The Lyman series and the the 2s two-photon
  decay spectrum. The main contribution to the Lyman series (thick solid line)
  comes from the Lyman-$\alpha$-transition (dashed line). The peak intensities for
  the first few transitions are $\Delta
  I_{\nu}=\pot{4.8}{-27},\,\pot{2.9}{-29},\,\pot{4.9}{-30}$ and
  $\pot{2.2}{-30}\,{\rm J\,m^{-2}\,s^{-1}\,Hz^{-1}\,sr^{-1}}$ for the
  Ly-$\alpha$, Ly-$\beta$, Ly-$\gamma$ and Ly-$\delta$, respectively.
  For comparison, the $2.725$~K blackbody is also shown (thin solid line).}
\label{fig:Lya_2sdecay}
\end{figure}

\subsection{The Lyman-$\alpha$ line}
As found by other authors, the Ly-$\alpha$ line leads to the
strongest distortion of the CMB spectrum and appears in the
Wien tail of the blackbody spectrum. 
In our computations it peaks at wavelength around 170~$\mu$m (see
Fig.~\ref{fig:Lya_2sdecay} and Table~\ref{tab:lines}) and is very similar in
shape and amplitude to the results obtained by \citet{Wong2005}. However, as
discussed below, we do not find any pre-recombination peak.

\begin{figure}
\centering \includegraphics[width=\columnwidth]{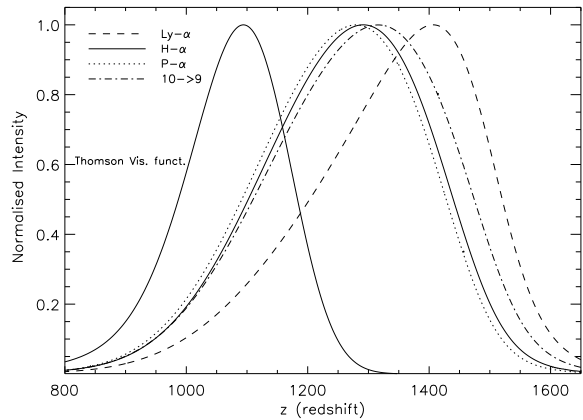}
\caption{Redshift of formation for different lines. We present the normalised
intensity for several transitions with $\Delta n = 1$ as a function of redshift
$z$. For comparison, we also show the Thomson visibility function
normalised to unity at the peak, showing that the redshift of formation of the
lines is higher than that of formation of the CMB fluctuations.}
\label{fig:z_of_formation}
\end{figure}

In Fig.~\ref{fig:z_of_formation} we present several transitions with $\Delta
n=1$ in redshift space.  The figure shows that the bulk of the Ly-$\alpha$
emission originates around $z \approx 1400$, as obtained by other authors
\citep[e.g.][]{Liubarskii1983}. 
Increasing $n$ one can see that the corresponding transition, due to the
decrease in the transition energy, stays longer in equilibrium with the photon
field. However, when the transition frequency close to the decoupling
redshift (i.e. the redshift at which the corresponding line starts appearing)
drops below the maximum of the CMB spectrum (here in terms of photon number,
$N_\nu\propto \nu^2/[\exp(h\nu/\kB\Tg)-1]$), i.e. $\nu_{\rm
max}\sim90.3\times(1+z)\,$GHz, then increasing $n$ further will shift the
redshift of formation for the corresponding line back to larger $z$ (see
Fig. \ref{fig:z_of_formation} transition $10\rightarrow 9$). For a more
quantitative estimate one also has to include the dependence of $A_{ij}$ on $n$
and eventually will end up looking at the net transition rates in more detail.

Here, another important point is that observations of these lines
permit us to explore an epoch of the Universe, where the optical depth
due to Thomson scattering is very high.  Note that the effects
discussed in \cite{Jose2005} also provide spectral information but,
due to the shape of the Thomson visibility function \citep{Suny70},
this information comes from lower redshifts ($z\sim 1100$).

\begin{figure}
\centering 
\includegraphics[width=\columnwidth]{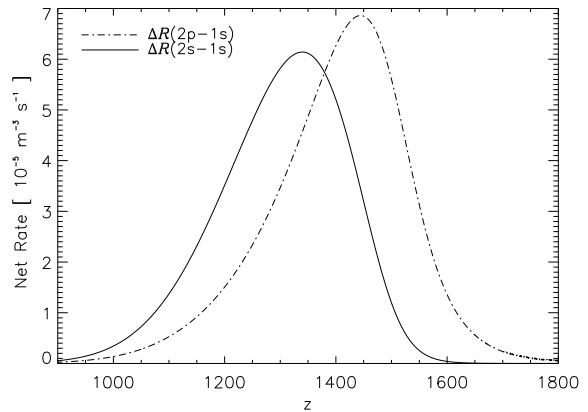}
\caption{Net rates for transitions from levels $n=2$ to the ground state.
  The rate $\Delta R({\rm 2p}-{\rm 1s})$ mimics the shape of the Ly-$\alpha$
  line (compare with Fig.~\ref{fig:z_of_formation}). 
  Note that the 2s two-photon decay rate dominates for redshifts
  smaller than $\sim 1400$ (the two rates are equal at $z\sim 1380$), i.e. 
  where the bulk of recombination takes place, and
  where all the contributions from higher transitions originate.}
\label{fig:net_rates_n2}
\end{figure}

Fig.~\ref{fig:net_rates_n2} shows the redshift dependence of the net
(radiative) rates of the Ly-$\alpha$ and 2s two-photon decay transitions.  As
it is well-known, the 2s$-$1s rate dominates during the epoch of
recombination, and provides the main channel through which recombination can
proceed \citep{ZeldKurtSun1968orig}.  On the other hand, escape
through the Ly-$\alpha$ line happens at earlier times (the two rates are equal
at $z\sim 1380$).  Note that due to the narrowness of the Ly-$\alpha$
line-profile, the shape of $\Delta R({\rm 2p}-{\rm 1s})$ mimics that of the
Ly-$\alpha$ contribution to the CMB spectral distortion (compare with
Fig.~\ref{fig:z_of_formation}).

\subsubsection{About the pre-recombination peak}
In earlier calculations of the recombination spectrum the existence of
a {\it pre-recombination peak} for the Lyman-$\alpha$ transition was
found \citep{RybickiDell94, Wong2005}.  However, in our calculations
we do not find any significant emission from any hydrogen line above
redshifts $z\sim 2000$, provided that the primordial radiation
field is described by a pure blackbody law.

With the assumption that there are {\it no intrinsic distortions} of the CMB
spectrum it is expected that at some high redshift the populations of all
neutral hydrogen states can be described using the Saha-relations, with tiny
deviations only coming from the disturbance of equilibrium due to the
expansion of the Universe. Clearly, in this case it is impossible to have any
uncompensated loops between different levels and hence no significant net
emission should occur \citep{Liubarskii1983}.
Our calculations show that, at redshifts $z \gtrsim 2000$, the net rate of the
2p$-$1s process becomes zero (within the numerical accuracy of the codes, which 
is much higher than the effects we are discussing here).

We found that here it is very important to obtain the solution for the
populations of levels with sufficiently high accuracy. In local thermodynamic
equilibrium (LTE) the Saha-relation yields\footnote{Note that in local
thermodynamic equilibrium $\Te\equiv \Tg$.}
\beal
\label{eq:Ni_Nj_Saha}
\left(\frac{N_j}{N_i}\right)^{\rm LTE}=\frac{g_j}{g_i}\,e^{h\nu_{ij}/\kB \Tg}
\end{align}
for the ratio of the population of the levels $i$ and $j$. This ratio
evidently is very sensitive to the value of the transition frequency
$\nu_{ij}$. Inserting this expression into Eq.  \eqref{eq:DRij} it is clear
that analytically one obtains {\it no} net emission, provided that $\nu_{ij} =
| \chi_i - \chi_j|/h$, with $\chi_i$ being the ionization potential of the
corresponding level.

However, in {\it numerical} calculations at sufficiently early epochs one will
obtain a solution for the population of each level, which is extremely close
to the {\it numerical LTE solution}.
But in fact, when one now computes $\frac{g_i}{g_j}\left(N_j/N_i\right)^{\rm
  LTE}_{\rm num}\,e^{-h\nu_{ij}/\kB \Tg}$ numerically even deviations of the
  order of $\Delta\left(N_j/N_i\right)^{\rm LTE}_{\rm
  num}/\left(N_j/N_i\right)^{\rm LTE}\sim 10^{-5}$ from the real LTE solution
  will lead to sizeable {\it false} emission, when deviations from LTE should
  be small.

This in principle can occur when using a slightly different value,
$\tilde{\nu}_{ij}=\nu_{ij}+\delta\nu_{ij}$ (for example from older
precalculated tables or using different combinations of the
natural constants for unit conversions) for the transition frequency
in the setup of the coupled system of ordinary differential equations,
while obtaining the output of the recombination spectrum with
$\nu_{ij}$. In other words, a false pre-recombination peak
appears when one is {\it not} comparing the obtained {\it numerical
solution} with the corresponding {\it numerical LTE solution}, which
in this setup should fulfill
\beal
\label{eq:Ni_Nj_Saha_num}
\left(\frac{N_j}{N_i}\right)^{\rm LTE}_{\rm num}=\frac{g_j}{g_i}\,e^{h(\nu_{ij}+\delta\nu_{ij})/\kB \Tg}
\Abst{.}
\end{align}
Inserting this expression into Eq.~\eqref{eq:DRij} one obtains
\beal
\label{eq:Prerec}
\Delta R_{ij}^{\rm num}(\nu)\propto 1-e^{h\delta\nu_{ij}/\kB \Tg}
\Abst{.}
\end{align}
Because of the large factor $\kappa=h\nu_{ij}/\kB T_0$ in the exponent even
for $\delta\nu_{ij}/\nu_{ij}\sim 10^{-6}$ this leads to notable {\it false}
emission at early epochs, while leaving the main recombination peak unchanged. 
As an example, for the Lyman-$\alpha$ transition ($\kappa\approx
43455$ and $\Delta\nu_\alpha/\nu_\alpha\sim 10^{-6}$) one obtains a
pre-recombination emission, which peaks at $z\sim 3100$ with $\Delta I\sim
\pot{6}{-26}\,{\rm J\,m^{-2}\,s^{-1}\,Hz^{-1}\,sr^{-1}}$, while the
recombination peak is roughly 10 times smaller!

One should mention that although at $z>2000$ there is a small difference in
the electron and photon temperature ($\Delta T/\Tg\sim 10^{-6}$) this cannot
lead to any pre-recombination emission:
due to the large entropy of the Universe ($N_{\rm b}/N_{\gamma}\sim 10^{-10}$)
one expects that even for a larger $\Delta T/\Tg$ the statistics of the
hydrogen levels is controlled by the photon field. This implies that in the
Boltzmann factors $\Tg$ appears instead of $\Te$. It is expected that even the
inclusion of collisional processes, which we neglected here, will not change
this fact. Similarly, the expected small change in the electron temperature
due to Compton heating a distorted photon field will not be of importance.
We therefore conclude that there should be no hydrogen pre-recombination
emission and that similarly one should not expect any pre-recombination
emission from helium.

\subsection{The two-photon decay spectrum}
In Figure~\ref{fig:Lya_2sdecay} we also show 
the contribution due to the 2s two-photon
decay. 
The peak of the emission is roughly at $\nu\sim 1460\,$GHz 
($\approx 205\,\mu$m) and partially fills the
gap between the Lyman and Balmer series (compare with 
Fig.~\ref{fig:main}). 
The amplitude of the 2s two-photon decay line is approximately 
$\pot{3.3}{-27}\,{\rm J\,m^{-2}\,s^{-1}\,Hz^{-1}\,sr^{-1}}$, which is 
only a factor $\sim 1.5$ times smaller than the peak value of the
Ly-$\alpha$ line. This ratio is compatible with the 
result found by \citet{Boschan1998}, while it differs from 
the result of \citet{Wong2005} by a factor of $2$, as a consequence of their
different normalisation of the profile function. 

Fig.~\ref{fig:net_rates_n2} also presents the 2s$-$1s net rate as a function
of redshift. Note that one can not directly infer from this figure the
position of the peak intensity of the contribution to the CMB spectral
distortion due to the 2s decay, because in this case the line profile (which
appears in the kernel of integration in Eq.~\ref{eq:I_2s1s}) is very broad.
Therefore, the resulting spectral distortion is much broader than any other
radiative line ($\Delta \nu/\nu \approx 0.8$, which is practically four times
larger than the value for Ly$\alpha$), and it peaks at higher frequencies
than in the simple extrapolation of the redshifted central frequency $\nu_{\rm
2s} \sim \nu_{\rm Ly\alpha}/2$ \citep{Wong2005}.

\begin{figure}
\centering 
\includegraphics[width=\columnwidth]{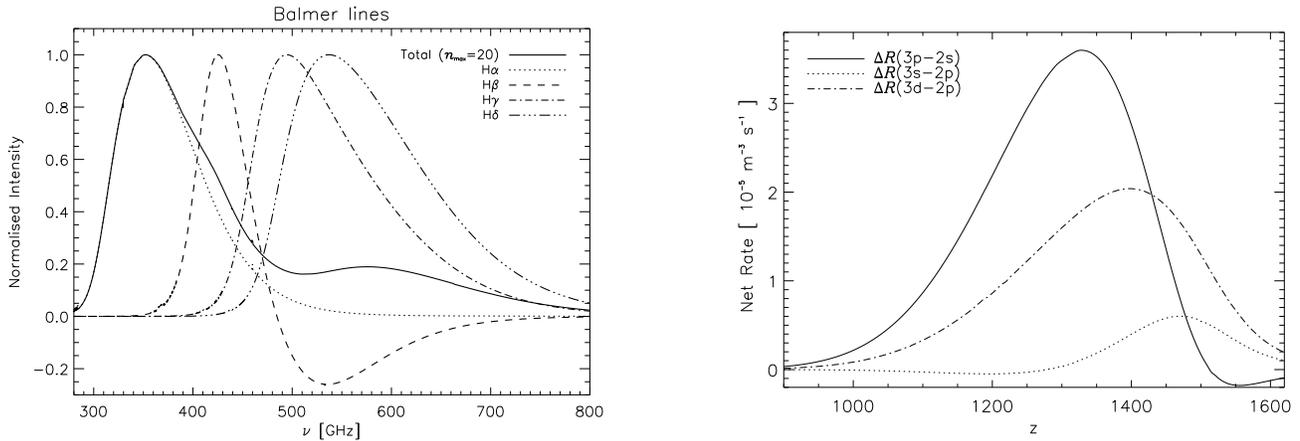}
\caption{Different contributions to the Balmer series. For convenience we
  normalised each of the contributions to unity at their maximum. The
  normalisation factors are $\Delta
  I_{\nu}=\pot{6.5}{-27},\,\pot{7.5}{-28},\,\pot{4.1}{-28}$ and 
  $\pot{2.5}{-28}\,{\rm J\,m^{-2}\,s^{-1}\,Hz^{-1}\,sr^{-1}}$ 
  for the H$\alpha$, H$\beta$, H$\gamma$ and H$\delta$, respectively.}
\label{fig:balmer}
\end{figure}

\begin{figure}
\centering 
\includegraphics[width=\columnwidth]{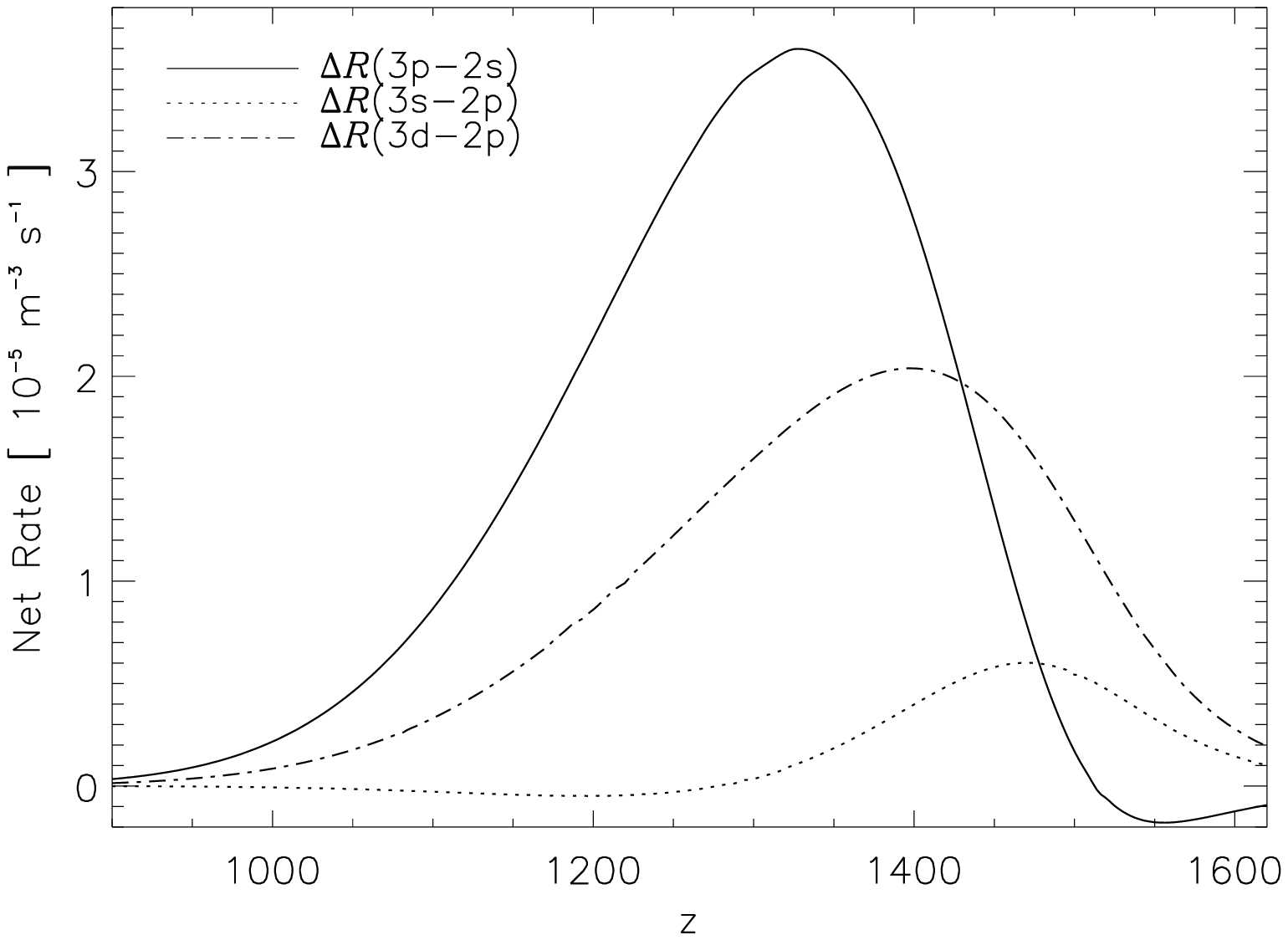}
\caption{Net rates for transitions from levels $n=3$ to levels $n=2$.
  This figure contains all transitions contributing to the
  H$\alpha$ line, namely $\Delta R({\rm 3s}-{\rm 2p})$, $\Delta R({\rm
  3p}-{\rm 2s})$ and $\Delta R({\rm 3d}-{\rm 2p})$, except the
  3p$-$1s. However, the contribution of the rate $\Delta R({\rm
  3p}-{\rm 1s})$ to the line is very small (it peaks at $z\sim 1470$
  with amplitude $\pot{4.6}{-7}$~m$^{-3}$ s$^{-1}$), so we can omit
  it. During the epoch of formation of the line, the fastest
  transition is 3p$-$2s, and therefore the population of the 3p level
  will be smaller than the expected equilibrium value (see
  Fig.~\ref{fig:nLTE_to_n}). }
\label{fig:net_rates_n3}
\end{figure}

\begin{figure*}
\centering 
\includegraphics[width=\columnwidth]{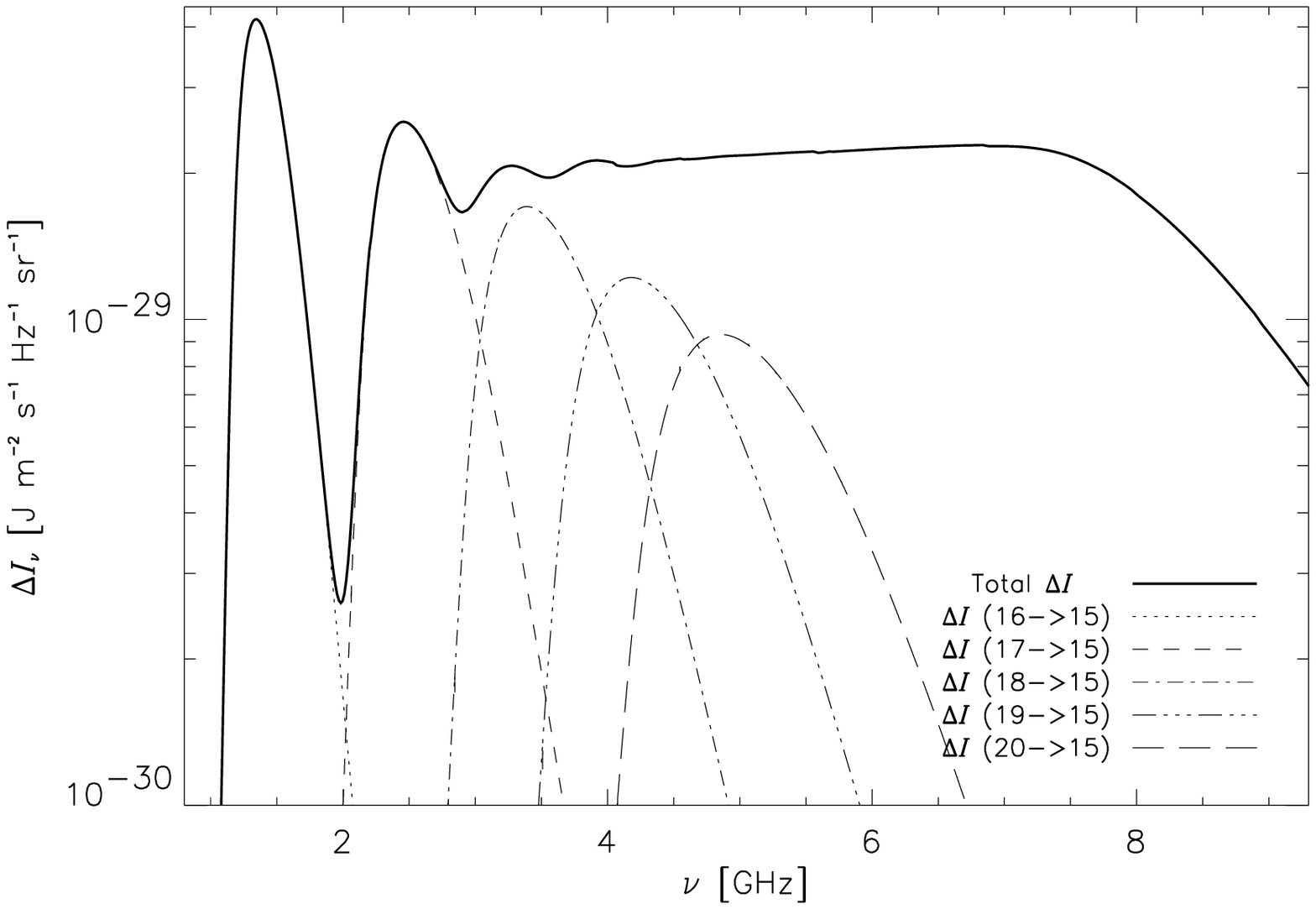}%
\includegraphics[width=\columnwidth]{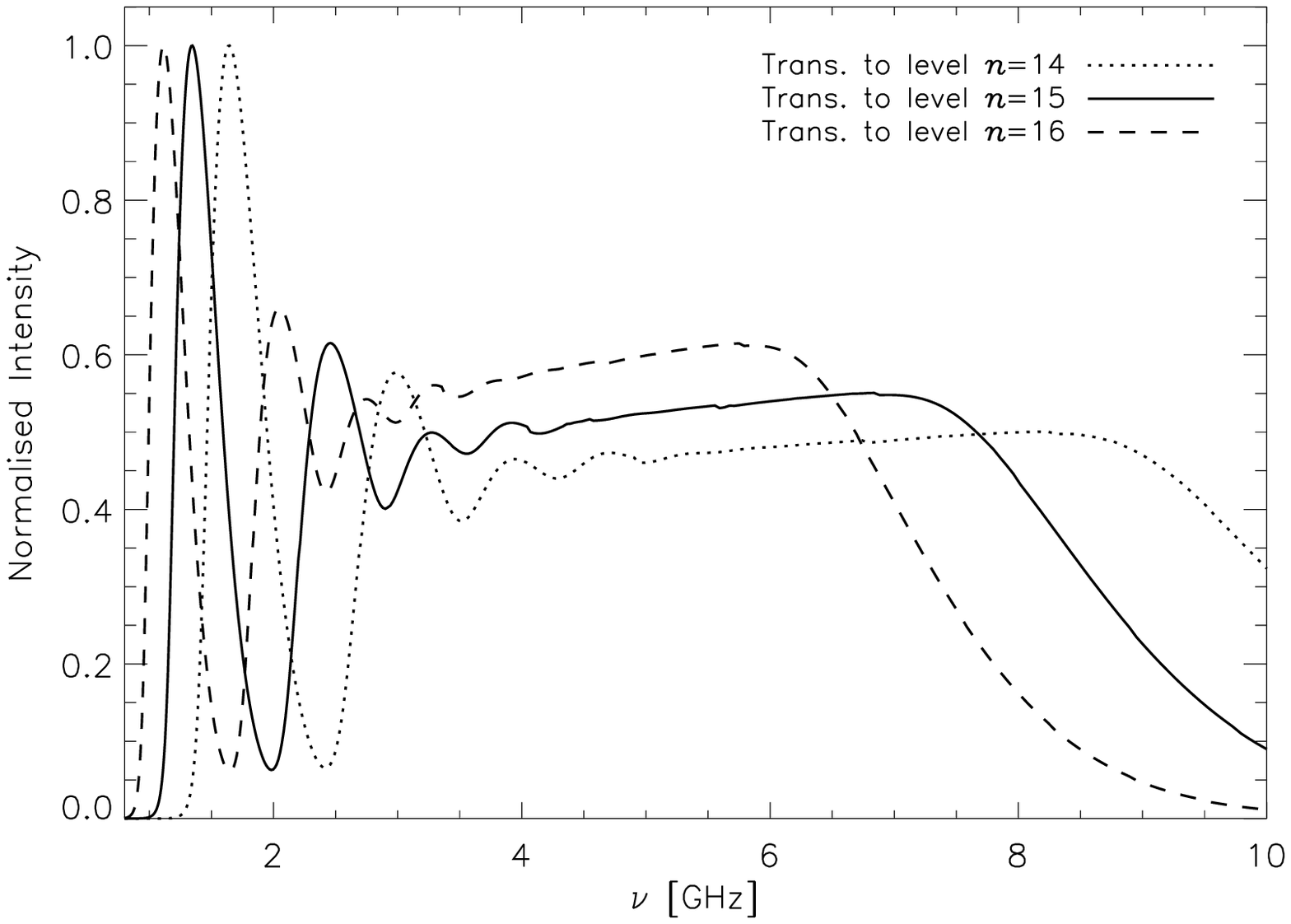}
\caption{Different contributions to a high-$n$ series. Left panel: we show the
contribution of all transitions to level $n=15$ for a hydrogen atom with
$\nmax = 30$. There is a strong peak produced by the corresponding 
$\Delta n=1$ transition, while the rest of transitions ($\Delta n \ge 2$) 
add-up to produce a (practically) featureless 
high-frequency wing of the series. 
Right panel: we show the net contribution of all transitions to levels
$n=14$, $15$ and $16$, in normalised intensity units. At frequencies below
a few GHz, the contribution of the $\Delta n=1$ transitions start
to overlap. }
\label{fig:high_transitions}
\end{figure*}

\subsection{The Balmer transitions and beyond}
\label{sec:balmer}
Fig.~\ref{fig:balmer} shows the different contributions to the Balmer
series using a linear scale in intensity.
The main contribution comes from the H$\alpha$ line. 
One can see that the contribution from H$\beta$ becomes negative 
in a certain frequency range. The positive wing of 
H$\beta$ slightly enhances the high-frequency
wing of the main feature at $350$~GHz, while the rest of the lines 
(H$\gamma$ and above) add up
to produce a secondary peak around $600$~GHz, completely compensating 
the negative part of H$\beta$.
Note that this result differs from that of \citet{2004AstL...30..509D}, where
their H$\beta$ and H$\gamma$ are both negative, and all the higher transitions
do not compensate them. Also in general the amplitude of our results 
is smaller by a factor of $\sim 10$.
We also mention that the results of \citet{RybickiDell93} for 
$\nmax =10$ contain negative features which are not found in our
computations (see Fig.~\ref{fig:convergence}). 

Fig.~\ref{fig:net_rates_n3} shows the net (radiative) rates for all
transitions connecting levels $n=3$ and $n=2$, and which contribute to
the H$\alpha$ line.  The fastest rate is associated with the 3p$-$2s
channel, as one would expect because the H$\alpha$ line forms around
$z \approx 1300$ (see Fig.~\ref{fig:z_of_formation}), where escape
through the 2s$-$1s channel is faster than through the 2p$-$1s one.

For higher series, one can perform a similar analysis.  For instance,
in the left panel of Fig.~\ref{fig:high_transitions} we show the first
contributions ($\Delta n \le 5$) to the series corresponding to level
$n=15$.  It is clear that for large values of $n$, the relative
distance between transitions with $\Delta n = 1$ and $\Delta n=2$
increases ($\Delta \nu/\nu \equiv [ \nu_{n,n+1} - \nu_{n,n+2}
]/\nu_{n,n+1} > 0.8$ for $n \ge 13$). Given that the typical width of
the lines is of the order of $0.3$ (see Table~\ref{tab:lines}), the
$\alpha$-transition is giving a separated peak, while the other
transitions of the series overlap and result in a nearly constant
emission plateau in the high-frequency wing. Although it is not shown
here, we have checked that the slope of this plateau is steeper than
the one obtained in the case of $\nsplit=2$. The co-added contribution
of all these high-frequency wings of all series is contributing to the
net slope at low frequencies in Fig.~\ref{fig:main}.

On the other hand, the peaks coming from the main ($\Delta n=1$)
transition of a given series cannot be separated from those of
other series for high values of $n$, because they start to overlap.
In the right panel of Fig.~\ref{fig:high_transitions} we show this
overlapping for the three series corresponding to $n=14$, $15$ and
$16$.  In this case, $\Delta \nu/\nu$ defined as $[ \nu_{n,n+1} -
\nu_{n+1,n+2} ] / \nu_{n,n+1}$ is a decreasing function with $n$, so
$\Delta \nu/\nu \le 0.2$ for $n\ge 13$, and thus the peaks start to
overlap and merge into a continuum. This effect can be seen in the
low-frequency part of Fig.~\ref{fig:main}.  Note that the co-added
contributions of all the high-frequency wings is giving the net
slope. If the populations of $l$-substates are not followed, the
transitions with $\Delta n>1$ will be larger, and thus we will have a
different behavior in the asymptotic slope in the sub-GHz range.

\subsection{Non-equilibrium effects of the angular momentum states 
on the electron fraction}

\begin{figure}
\centering
\includegraphics[width=\columnwidth]{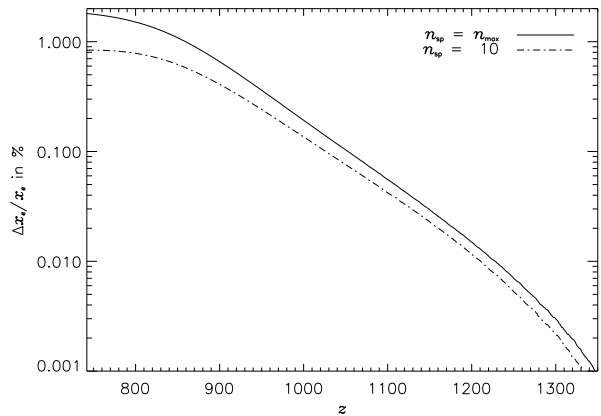}
\caption{Relative change in the free electron fraction ($x_{\rm e} =
N_e/N_H$).  We consider the 25-shell atom and show the relative difference in
$x_e$ with respect to the standard computation ($\Delta x_{\rm e}/x_{\rm e}
\equiv [x_{\rm e} -x_{\rm e}({\nsplit=2})]/ x_{\rm e}({\nsplit=2})$) for two
cases: full computation ($\nsplit=\nmax$) and $\nsplit=10$. }
\label{fig:xe}
\end{figure}

One of the important variables in our calculations is the electron
fraction, $x_{\rm e}= N_e/N_H$, as a function of redshift $z$. Here we
want to quantify the effects on the electron fraction.
To do so, we compare the results obtained for the 25-shell atom in the full
calculation ($\nsplit=25$) with those for standard computation ($\nsplit=2$).
Fig.~\ref{fig:xe} shows the relative change of the free electron fraction due
to the inclusion of angular momentum substates. In the standard computation
the residual electron fraction at low redshifts ($z=800$) is smaller by $\sim 1.5$\%
than in the case of full $l$-follow-up, implying that in this case 
the process of recombination is slightly slower. Note that for $\nsplit=10$ 
one obtains approximately half of the effect.

The net effect during the epoch of recombination \citep[$z_{\rm dec}=1089\pm
1$, ][]{WMAP_params} is smaller than 0.1\%. Therefore one would only expect a
small impact on the angular power spectrum of the CMB.
To quantify this effect we have modified the {\sc Cmbfast}
\citep{cmbfast} code, to incorporate the recombination history
obtained in our computations. We note that this recombination history is
computed inside {\sc Cmbfast} using the {\sc Recfast}
\citep{Seager1999ApJ} code, which is based on calculations for a
$300$-level hydrogen atom in which SE among the $l$ sublevels was
assumed \citep{Seager2000ApJS}. 
Our computation makes the assumption that {\it the relative change of
the electron fraction, as obtained for the 25-shell atom, will not vary
significantly when including more shells.} Although the inclusion of more
shells in our calculations should still lead to a decrease of the residual
electron fraction at low redshifts (one would expect to converge for $\nmax\sim
50$), we found that the above 
assumption is justified reasonably well. We
will further discuss this point in a future work. 

\begin{figure}
\centering 
\includegraphics[width=\columnwidth]{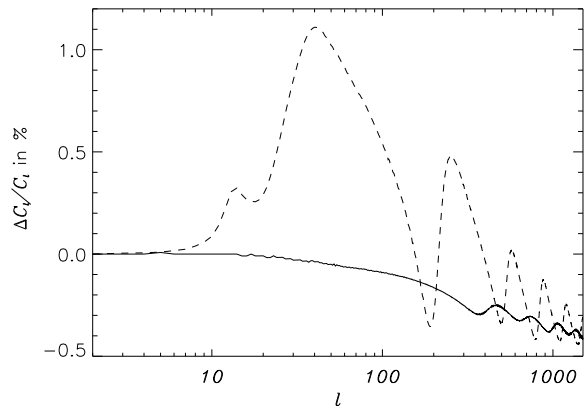}
\caption{Relative change in the temperature (TT, solid line) 
and polarization (TE, dashed line) angular power spectra due
to the inclusion of detailed follow-up of the angular momentum sub-states, 
for the WMAP best-fit cosmological model. }
\label{fig:cmbfast}
\end{figure}

Fig.~\ref{fig:cmbfast} shows that the relative difference of the $\cl$ is
smaller than $\sim 0.5$\% ($\sim 1.1$\%) for the TT (TE) angular power
spectrum.
Following \citet{Seager1999ApJ} we can model the results by multiplying the
recombination/ionization rates by a `fudge factor'. Our results for $x_{\rm
e}$ are reproduced with an extra factor $0.967$, i.e. the fudge factor inside
{\sc Recfast} should be $1.10$ instead of $1.14$.
%

Finally, we would like to mention that there are other effects, which
may produce small changes in the $\cl$'s. For example,
\citet{Chluba2006} considered the effect of stimulated emission due to
the presence of soft CMB photons on the total 2s-decay rate. In
addition, \citet{Leung2004} discussed the effect of a modification of
the adiabatic index in the matter equation of state as recombination
proceeds and \citet{Dubrovich2005} included corrections due to the
two-photon decay of higher hydrogen levels and helium. All the
aforementioned processes lead to changes of the order of 1\% and even
cancel eachother partially, leaving a lot of space for future work and
discussion.

\section{Discussion}
\label{sec:discusion}
The spectrum of the background radiation of the Universe has a minimum
in the vicinity of $30-40$~cm \citep{LongairSunyaev72}.  It is very
impressive that the brightness temperature in the recombinational line
component has a maximum relative to the background radiation spectrum
precisely in this spectral band (see
Fig.~\ref{fig:relative_distortion}), and reaches values of the order
of $10^{-7}$.  This is a consequence of the fact that the asymptotic
slope of the spectral distortions to the blackbody (which in our case
is of the order of $\approx +0.8$) is decaying slower with frequency
than the Rayleigh-Jeans part of the CMB (slope $+2$). Note that the
turn-over at around 0.5~GHz is due to the lack of diffusion from high
levels in our calculation with $\nmax=30$.

It is extremely difficult to separate an additional continuum component with
quasi-flat spectrum from the foreground sources.  However, the lines with
$\Delta n = 1$ will still influence the shape of the integral curve, and the
contribution from recombination is modulated in such a way that no foreground
source is able to mimic it. Obviously, one of the most interesting spectral
regions corresponds to those frequencies just above $1420$~MHz ($21$~cm line),
where the contribution of synchrotron emission of relativistic electrons of
radio and ordinary galaxies \citep[e.g.][]{LongairSunyaev72,1993ebr..proc..223L}
and free-free emission from our Galaxy are not expected to have a quasi-periodic
dependence on the wavelength (for a more detailed discussion of
foregrounds for 21~cm studies, see \citet{2003MNRAS.346..871O,
2005AAS...207.3304M}).

\begin{figure}
\centering 
\includegraphics[width=\columnwidth]{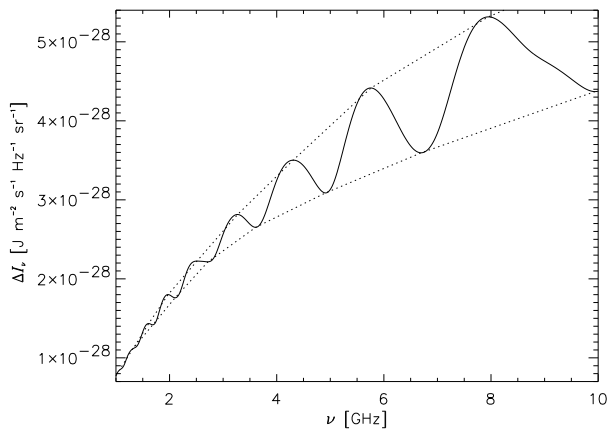}
\caption{A detail of the hydrogen spectrum in the frequency domain
between 1~GHz and 10~GHz. We present in linear scale the amplitude of
the features in the region of the spectrum were we expect that detection
should be more feasible. Dotted lines are connecting the maxima and
minima of the features, so one can read directly their contrast. }
\label{fig:zoom}
\end{figure}

At low frequencies, the amplitude of this frequency
modulation is not strong, but it might become measurable by future
experiments \citep[see e.g.][]{Fixsen2002}. Fig.~\ref{fig:zoom} shows
that the contrast of these features in the frequency range between 1
and 10~GHz is between 5\% and 30\%.
Even though the amplitude of these features may still increase
when considering larger values of $\nmax$, the contrast should not change
significantly.
Note that the Balmer, Paschen and Brackett series have a higher
contrast and they are found in the frequency band where the
contamination is smaller, but they will be much more difficult to
observe because they lie two orders of magnitude below in relative
intensity (see Fig.~\ref{fig:relative_distortion}).
We would like to mention here that there have been observational
efforts to measure the CMB spectrum at centimeter wavelengths
\citep{2004ApJS..154..493K, 2004ApJ...612...86F}, searching for
signatures of spectral distortions from early energy injection in the
spectrum \citep{Suny70b, 1975SvA....18..691I}.

Although the Lyman-$\alpha$ distortion is the strongest feature in the
spectrum (relative distortion in the peak is of order unity), the
Cosmic Infrared Background (CIB) in this region is dominant, making its
observation extremely difficult.
However, the sensitivity of the detectors is improving a lot, so these
observations could become feasible in the future.
For example, the recent measurements of Spitzer \citep{2006astro.ph..3208D} have
shown that at 160~$\mu$m, practically three quarters of the Mid-Infrared
Background (MIB) can be already resolved into luminous infrared galaxies at
$z\sim 1$. If the MIB could be completely resolved, one might be able to remove
the source contamination in a given piece of sky in order to measure the
spectral distortions discussed here.

\begin{figure}
\centering 
\includegraphics[angle=90,width=\columnwidth]{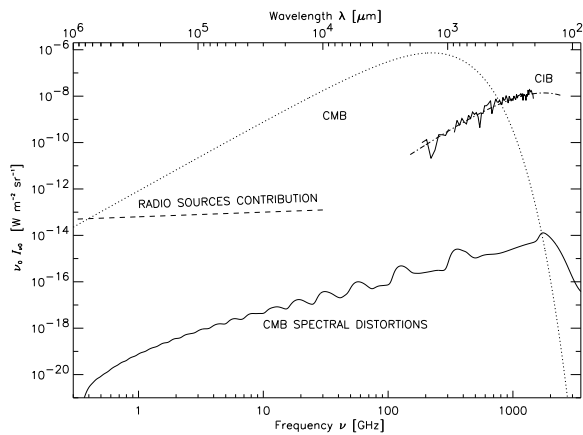}
\caption{Spectral distortions and foregrounds. We compare the result of
Fig.~\ref{fig:main} with the pure blackbody spectrum of $2.725$~K (dotted line);
with the emission of the Cosmic Infrared Background derived by
\citet{1998ApJ...508..123F} (dot-dashed line) and \citet{2000A&A...354..247L}
(solid line between 200$\mu$m and 1.2mm); and with the estimates of the
contribution of extragalactic radio sources from \citet{1993ebr..proc..223L}
(note that this refers to the extragalactic sources with an average
spectral index of $-0.8$ in flux, and which roughly scales as $6$~K $\times
(\nu/178~{\rm MHz})^{-2.8}$; this is why in these coordinates, their contribution
seems practically flat). }
\label{fig:cib}
\end{figure}

For illustration, Fig.~\ref{fig:cib} shows the amplitude of the
spectral distortions compared to the determination of the CIB obtained
with COBE FIRAS data \citep{1998ApJ...508..123F}, and the result of
\citet{2000A&A...354..247L} in the Lockman Hole using the WHAM
H$\alpha$ survey and the Leiden/Dwingeloo HI data.
We note that, in addition to the CIB signal, there will be also a
strong contamination from our Galaxy \citep[see the discussion
in][]{Wong2005}.

We would like to stress that all the results of this work were obtained
within the standard framework which has been adopted by many other authors
working on this topic (see the references in the introduction).  The main
difference with respect to these earlier works is the detailed treatment of the
$l$-substates. However, it is clear that one has to evaluate the impact of other
physics which has been neglected so far.
As an illustration, we will discuss three additional physical
processes which have not been included here and which could modify the results.

The first effect is connected with collisions.  Although the collisional rates
are smaller than the radiative rates during the epoch of recombination
\citep{Seager2000ApJS}, they could produce a redistribution of electrons if we
consider states which are not directly connected via dipolar electric
transitions (e.g. collisions can produce transitions of the type
$(n,l)\rightarrow(n,l')$ also for $\Delta l\neq 1$). In principle, this effect
should be important for shells with large principal quantum number $n$, and
would tend to restore SE, thus restoring the $(2l+1)$-ratios of the populations
of the levels within a given shell.  In other words, one should eventually
recover the non-splitting solution for sufficiently high shells, and the
asymptotic slope of $+0.35$ found by \citet{Kholu2005} might be reached.
One can do a simple estimate of the effect by using the collisional
rates described by Eq.~43 of \citet{1964MNRAS.127..165P}.  In our
problem, the collisional transition probability $q_{n l} N_{n l} N_e$
becomes of the order of the radiative transition probability $A_{nl}$
for $n \ga 25-30$ at the redshift of formation of the lines ($z\approx
1300$).  Thus, if one wants to include a higher number of shells in the
computations, the effect of collisions 
on the mixing of the populations in different $l$-sublevels 
has to be taken into account.

The second effect is the free-free absorption \citep[see
e.g.][]{1975SvA....18..691I, 2004AstL...30..509D}, which at low frequencies 
will tend to restore the blackbody with $\Tg=\Te$ 
and thus will erase the features in the spectrum. 
However, this effect will be only important for high shells,
so our computation up to $\nmax=30$ will not change significantly.

Finally, also the inclusion of additional radiative channels (stimulated
two-photon decay from higher levels, forbidden transitions, i.e. both electric
and magnetic quadrupole transitions, or magnetic dipole transitions) may have
some impact on the computed spectral distortions.
However, we would expect that the inclusion of these new channels
will not subtantially modify the spectral distortions discussed here.
The reason for this is that all the sub-levels with $n>2$ can connect
to lower states via permitted radiative transitions having escape
probability very close to unity. Therefore, the inclusion of additional
"slow channels" should only modify our results at the level of a few
percent.
We have already checked that the inclusion of stimulated 2s two-photon decay
\citep{Chluba2006} and some additional two-photon decay channels for higher
hydrogenic levels \citep[following the prescription in][]{Dubrovich2005}, leads
to differences with respect to our computations of the order of 1\%, in both the
recombination spectrum and the electron fraction.

Certainly one should try to develope an approach including all the
aforementioned processes stimultaneously, but this shall be left for a future
work.

\section{Conclusions}

In this paper, we have computed the spectral distortions of the CMB frequency
spectrum originating from the epoch of cosmological hydrogen
recombination, by following the evolution of the populations of the levels of
a hydrogen atom with $\nmax = 30$ in the redshift range from $z=3500$ down to
$z=500$.
Our computation follows in detail the evolution of 
all angular momentum sub-states. This permits us to describe the
non-equilibrium effects which appear during the process of recombination
in the populations of the sub-levels.

The populations of the levels are found to differ (with respect to the
standard computation without detailed follow-up of $l$-states) by more
than 20\% for low redshifts ($z \la 800$). In addition, the amplitude
of the lines is also affected. Following all the angular momentum 
substates, the relative ratio of the different lines within a given series
increases. This can be seen, for example, in the case of the Balmer
series, where the amplitude of the $\alpha$-transition ($\Delta n =1$)
increases while the rest of transitions decrease.  This shows that
the simplifying assumption of statistical equilibrium within 
a given shell leads to an overall underestimation of the net emission at
high frequencies.

We have also quantified the effects on the residual electron fraction, 
showing that at low redshifts ($z \approx 800$) the $x_e $ is increased by 1.5\%. 
The impact of this change on the angular power spectrum is 
small, producing differences of the order of $1\%$ in
the $\cl$'s.

Finally, our computations have shown that, if the primordial
radiation field is assumed to be a pure blackbody spectrum, then 
there is no significant emission from any line for redshifts greater
than $z=2000$, so a ``pre-recombination'' peak does not exists.

\section*{Acknowledgments}
We acknowledge use of the {\sc Cmbfast} software package \citep{cmbfast}.  JAR-M
acknowledges the hospitality of the MPA during several visits, and thanks
Andr\'es Asensio Ramos for useful discussions.  JC thanks the IAC for
hospitality during his visit in March 2006.  We also thank our referee,
Douglas Scott, for his useful remarks and his interest in the paper.


\label{lastpage}

\end{document}